\documentclass[aps,prd,superscriptaddress,showpacs,amsmath,amssymb,amsfonts,nofootinbib,showpacs,reprint]{revtex4-1}

\pdfoutput=1

\usepackage{graphicx}
\usepackage{color}
\usepackage{slashed}
\usepackage{tabularx}
\usepackage{hyperref}
\usepackage{tkz-graph}
\usepackage{ifthen}
\usepackage{dsfont}

\def\simge{\mathrel{
     \rlap{\raise 0.511ex \hbox{$>$}}{\lower 0.511ex \hbox{$\sim$}}}}
\def\simle{\mathrel{
     \rlap{\raise 0.511ex \hbox{$<$}}{\lower 0.511ex \hbox{$\sim$}}}}
\def\be{\begin{equation}}
\def\ee{\end{equation}}
\def\bea{\begin{eqnarray}}
\def\eea{\end{eqnarray}}

\newcommand{\calz}{{\cal Z}}

\newcommand{\Tr}{{\rm Tr}\,}

\newcommand{\expo}[1]{\text{exp}\left\{ #1 \right\}}

\newcommand{\eq}[1]{\begin{eqnarray}#1\end{eqnarray}}  
\newcommand{\eqa}[1]{\begin{align}#1\end{align}}
\newcommand{\eqal}[2]{\begin{align}\label{#2}#1\end{align}}

\newcommand{\pathint}[1]{\int \mathcal{D} #1 \, }
\newcommand{\beq}{\begin{eqnarray}}
\newcommand{\eeq}{\end{eqnarray}}

\newcolumntype{L}[1]{>{\raggedright\arraybackslash}p{#1}} 
\newcolumntype{C}[1]{>{\centering\arraybackslash}p{#1}} 
\newcolumntype{R}[1]{>{\raggedleft\arraybackslash}p{#1}} 

\graphicspath{{.}{figures/}{mma/}{figures}}

\definecolor{blue}{rgb}{0,0,1}

\definecolor{green}{rgb}{0,1,0}

\definecolor{red}{rgb}{1,0,0}

\begin{document}

\title{A density of states approach to the hexagonal Hubbard model at finite density}

\newcommand{\JLU}{Institut f\"ur Theoretische Physik, Justus-Liebig-Universit\"at, Heinrich-Buff-Ring 16, 35392 Giessen, Germany}
\newcommand{\UL}{School of Mathematics, University of Leeds, Leeds, LS2 9JT, UK}

\author{Michael K\"orner}\affiliation{\JLU}
\author{Kurt Langfeld}\affiliation{\UL}
\author{Dominik Smith}\affiliation{\JLU}
\author{Lorenz von Smekal}\affiliation{\JLU}

\begin{abstract}
We apply the Linear Logarithmic Relaxation (LLR) method, which generalizes the Wang-Landau algorithm to quantum systems
with continuous degrees of freedom, to the fermionic Hubbard model with repulsive interactions on the honeycomb lattice. 
We compute the generalized density of states of the average Hubbard field and divise two reconstruction schemes to
extract physical observables from this result. By computing the particle density as
a function of chemical potential we assess the utility of LLR in dealing with the sign problem of this model, which
arises away from half filling. We show that the relative advantage over brute-force reweighting grows as the 
interaction strength is increased and discuss possible future improvements.
\end{abstract}

\maketitle

\section{Introduction}\label{sec:intro}

Monte Carlo simulations based on path-integral
quantization are a powerful and widely used tool for the
study of strongly coupled quantum systems. They are applied
in many different areas of physics, ranging from
high-energy physics, where they are employed e.g. to
study the phase diagram and particle spectrum of Quantum Chromodynamics (QCD), 
to condensed matter physics, where they are used to study
strongly correlated electron systems. Quite often, they are the
only way to obtain reliable information from first principles. 
Unfortunately, their applicability is restricted to a very special
class of systems, namely those where the path integral
exhibits a positive-definite measure which can be interpreted
as a probability density. 
This excludes most fermionic systems
away from half filling (unless the complex parts of the measure 
cancel exactly due to an anti-unitary symmetry)
as well as quantum systems which evolve in real (as opposed to Euclidean) time.
This restriction is known as the \emph{sign problem} and is a long-standing problem
of numerical physics. 

In principle, brute-force reweighting techniques can be used
to extract information about systems with charge imbalance from 
simulations of a corresponding system at charge neutrality.
These are plagued by a severe signal-to-noise ratio problem
however, originating from a loss of overlap between 
the ensembles at zero and non-zero charge density when the thermodynamic 
limit is approached, and typically fail well below $\mu/T \approx 1$ except on very
small systems. While a theorem by Troyer and Wiese states that the sign problem is 
a NP-hard problem in a generic spin-glass system \cite{Troyer:2004ge} 
(making the discovery of a complete universal solution to all sign problems unlikely), 
many attempts have been made to construct specialized solutions for specific systems 
(typically by introducing a set of dual variables), or improved 
general approaches which outperform reweighting, with some success. Most notably,
in QCD, Taylor expansions of the partition function with respect to $\mu/T$ have now been pushed
to $\mu/T \approx 2$ or $3$ \cite{Bazavov:2017dus}. 

A promising, quite different, idea to deal both with ergodicity problems
in Monte Carlo simulations of systems close to a first order phase transition, 
and overlap problems resulting from a non-positive probability measure, is to
use non-Markovian random walk simulations, which do not rely on
importance sampling with respect to a positive Gibbs factor. A particularly
interesting class of algorithms use the inverse density-of-states as a measure for 
updating configurations. This measure is positive (semi-)definite by definition, 
and produces a random walk which efficiently samples configuration space even in 
``deprived'' regions with very low probabilistic measure. Prominent
examples are the multicanonical algorithm by Berg and Neuhaus \cite{Berg:1992qua} 
and the Wang-Landau approach \cite{PhysRevLett.86.2050}, which both were
designed for theories with discrete degrees of freedom. 

The \emph{Linear Logarithmic Relaxation} (``LLR'') method, first
described in Ref. \cite{Langfeld:2012ah}, generalizes this idea
to quantum systems with continuous degrees of freedom.
The goal of LLR is to estimate the slope $a(X)=\frac{d}{dX} \ln(\rho(X))$, where $X$ is 
some observable and $\rho(X)$ is the ``generalized density of states'' (gDOS). 
Once $a(X)$ is obtained, $\rho(X)$ can be reconstructed up to a 
multiplicative factor by numerical integration and can be used to compute thermodynamic observables. 
A crucial property of LLR is that it features exponential error suppression: 
The estimate for $a(X)$, and by extension of $\ln(\rho(X))$, can be obtained with roughly the same statistical error,
independent of the exact value of $X$, even if $X$ is from a region of overall low weight. 

In recent years, LLR was successfully applied to obtain 
$\rho(E)$ in SU(2) and U(1) \cite{Langfeld:2012ah} as well as SU(3) \cite{Gattringer:2016kco}
gauge theories and was shown to be effective at dealing with ergodicity issues
arising at first-order phase transitions in U(1) gauge theory \cite{Langfeld:2015fua}
and the $q=20$ state Potts model \cite{Langfeld:2016kty}. 
In Ref. \cite{Langfeld:2013xbf} LLR was applied to obtain the Polyakov loop
distribution for two-color QCD (which has no sign problem) with heavy quarks at finite densities. 
To deal with the sign problem, one needs to compute the gDOS for the imaginary part
of the Euclidean action $\rho(S_I)$, or some related observable. This was achieved using 
LLR for a Z$_3$ spin model 
at finite charge density \cite{Langfeld:2014nta} and for QCD in the heavy-dense limit \cite{Garron:2016noc}.
To date, LLR has never been applied to a sign problem of a system with fully dynamical fermions however. 

In this work, we apply LLR to the fermionic Hubbard model on the honeycomb lattice away
from half filling within a Hybrid Monte Carlo framework. Despite its simplicity, the Hubbard model, which describes
fermionic quasi particles with contact interactions, continues to be of profound 
interest, as it remains the quintessential example of an interacting fermion system,
and can qualitatively describe many non-perturbative phenomena such as dynamical 
mass-gap generation or superconductivity. On the honeycomb lattice, this model exhibits a second order phase transition in the universality class of the chiral Gross-Neveu model in 2+1 dimensions
\cite{Assaad:2013xua,PhysRevX.6.011029,PhysRevB.91.165108,PhysRevB.90.085146}.
With its relativistic dispersion relation for the low-energy excitations in the Dirac-cone region it therefore also provides a convenient lattice regularization, with minimal doubling, of relativistic theories for chiral fermions with local four-fermi interactions such as the Gross-Neveu or Nambu-Jona-Lasinio models which are of continued interest in searches for inhomogeneous phases \cite{Lenz:2020bxk} as predicted also for the QCD phase diagram, mainly from mean-field studies of the NJL model \cite{Buballa:2014tba}. Extended versions of the hexagonal Hubard model, which include long-range interactions, are also used to realistically describe the physics of both mono- and bilayer graphene \cite{PhysRevLett.106.236805,PhysRevX.8.031087}.

Using LLR, here we compute the 
gDOS for the average of a real-valued auxiliary field, which is 
introduced in Hybrid Monte Carlo simulations to transmit inter-electron interactions.
 We demonstrate that this result can be 
used to reconstruct the fermion density as a function of chemical potential. We show that,
in its present form, LLR can probe much further into the finite density regime than
standard reweighting, that the relative advantage of LLR grows as the interaction strength is increased, and 
argue that future improvements might put the van Hove singularity in the single-particle bands within reach.

This paper is structured as follows: We start in Sec.~\ref{sec:setup} by introducing the basic lattice setup and illustrating the sign problem away from half filling. 
Subsequently, we introduce the generalized density of states of the average Hubbard field $\rho(s)$ in Sec.~\ref{sec:doshubbard}.
In Sec.~\ref{sec:reconstruct}, we discuss the reconstruction of the particle density $n$ from $\rho(s)$. We describe two different reconstruction schemes, whereby $n(\mu)$ is obtained from both the canonical and grand-canonical partition functions.
As a benchmark, we apply both schemes to test data obtained for the non-interacting tight-binding theory. Full LLR calculations of the interacting theory, including additional numerical details, are then presented in Sec. \ref{sec:llr_results}. We summarize and conclude in Sec.~\ref{sec:outlook}.

\section{Lattice setup and the sign problem}\label{sec:setup}

We consider the repulsive Hubbard model on the hexagonal (honeycomb) lattice with fermionic creation operators $\hat c_x^\dagger \equiv (\hat c_{x,\uparrow}^\dagger,\hat c_{x,\downarrow}^\dagger) $ for two spin components at site $x$, which is defined by the effective Hamiltonian for the grand canonical ensemble:
\begin{align} 
\label{eq:tightbinding}
\hat{\mathcal{H}} =& - \kappa \sum_{\left \langle x, y \right \rangle} ( \hat{c}^{\dagger}_{x} \hat{c}_{y} + \text{h.c.})\nonumber\\
& + \sum_{x} \Big( m_s \, \hat{c}^{\dagger}_{x} \sigma_3 \hat{c}_{x}  + \frac{U}{2}\,
\hat{\rho}_x^2 - \mu\, \hat{\rho}_x  \Big)~ .
\end{align}
Here $\kappa$ is the hopping parameter, which quantifies the energy cost of fermionic quasi-particles propagating between nearest-neighbor sites.  Its phenomenological value in the tight-binding description of the electronic properties of graphene on a substrate is $\kappa \approx 2.7 \textrm{eV}$. In general, it is typically used to set the energy scale in the Hubbard model. We work in a natural system of units and therefore express all dimensionful quantities in terms of $\kappa$, which effectively corresponds to setting $\kappa\equiv 1$. The sum over $\langle x, y\rangle$ sums all independent pairs of 
nearest-neighbors, $m_s$ is the sublattice-staggered mass term (with alternating sign on the two triangular sublattices) for explicit sublattice-symmetry breaking 
with spin-density-wave order. The  chemical potential
  $\mu $ couples to the charge operator $\hat{\rho}_x =\hat{c}^{\dagger}_{x} \hat{c}_{x} -1 $ and controls the charge-carrier density. Experimentally this is achieved through chemical doping \cite{PhysRevB.94.081403} or electrolytic gate voltages \cite{PhysRevLett.105.256805}, for example. 
The constant $U$ controls the interaction strength, which is positive in the repulsive Hubbard model. 
The creation and annihilation operators satisfy the fermionic anticommutation relations $\{ \hat{c}_{x}, \hat{c}^{\dagger}_{y} \}= \delta_{x,y} \, \mathds{1}$. 
Lattice simulations of (\ref{eq:tightbinding}) using Hybrid Monte-Carlo by now have a long history already
\cite{Brower:2012zd,Buividovich:2012nx,Ulybyshev:2013swa,Smith:2014tha,Smith:2013pxa,Smith:2014vta,PhysRevB.96.165411,Buividovich:2016tgo,PhysRevB.93.155106, Korner:2017qhf,Beyl:2017kwp,Buividovich:2018hubb,Buividovich:2018crq,Wynen:2018ryx,Krieg:2018pqh,Smith:2019fbh}, we thus summarize only the essential steps here.\footnote{
In particular we omit the discussion of fermionic coherent states and the partial particle-hole transformation that is applied. 
These and other details can be found e.g.~in Refs.~\cite{Smith:2014tha,Buividovich:2018hubb,Smith:2019fbh}.} 

To derive the functional integral representation of the partition function at inverse temperature $\beta = 1/T$, we first write the exponential in terms of $N_t$ identical factors and split the Hamiltonian into the free tight-binding part plus interactions, $\hat{\mathcal H} = \hat{\mathcal{H}}_{\text{0}} + \hat{\mathcal{H}}_{\text{int}}$. A symmetric Suzuki-Trotter 
decomposition of each of the factors then yields 
\begin{eqnarray}
\calz &=& \Tr \left( e^{-\beta \hat{\mathcal{H}} } \right) \notag\\ &=&
\Tr \left( e^{-\delta \hat{\mathcal{H}}_{0}} e^{-\delta \hat{\mathcal{H}}_{\text{int}}} e^{-\delta \hat{\mathcal{H}}_{0}} \dots \right) + O(\delta^2 ).
\label{eq:Trotter}
\end{eqnarray}
This introduces a finite step size $\delta = \beta/N_t $ in Euclidean time and a discretization error of $O(\delta^2 )$. 

As we will see shortly, it is convenient to include the chemical-potential term in the definition of $ \hat{\mathcal{H}}_{\text{int}}$ here, i.e.~defining
\begin{equation}
  \hat{\mathcal{H}}_{\text{int}} \equiv  \sum_{x} \Big( \frac{U}{2}\,
\hat{\rho}_x^2  - \mu\, \hat{\rho}_x \Big)~.
\end{equation}
The four-fermion interaction in $\hat{\mathcal{H}}_{\text{int}}$ is then converted to bilinears by Hubbard-Stratonovich transformation,
\begin{eqnarray}
\label{continuous_HS_imag}
  e^{-\delta \hat{\mathcal{H}}_{\text{int}}}
\cong \pathint{\phi} \,
  e^{- \frac{\delta}{2U} \underset{x}{\sum} \phi_x^2 }\,  e^{-i \delta \underset{x}{\sum} (\phi_x + i \mu ) \hat \rho_x }\,, 
\end{eqnarray}
whereby the auxiliary (``Hubbard-Coulomb'') field $\phi_{x,t}$ is introduced. 
Finally, the trace over the fermionic operators is performed by integrating the fermionic coherent states  \cite{Buividovich:2018hubb}, which yields
\eqa{\label{eq:partfunc}
\calz  =\pathint{\phi} \det{\left[ M(\phi , \mu) M^\dagger(\phi,-\mu ) \right]} \,
\expo{-\frac{\delta}{2U} \underset{x,t} \sum 
\phi_{x,t}^2}.} 
Different versions of the fermion matrix $M(\phi)$ have been used in the past which are either equivalent or at least yield the same continuum limit. In this work we use
\eqa{
M(\phi,\mu) _{(x,t),(x',t')} = \, & \delta_{xx'} \exp\{ i
\delta \, (\phi_{x,t}  + i \mu) \}   \delta_{tt'}
\nonumber \\
- \, &\Bigl( \delta _{xx'} -  \delta h_{xx'} \Bigr) 
\delta_{t+1,t'}   ,
\label{eq:fermionmat} \\
h_{xx'}  = \, &  \delta_{xx'}  \,  m_s  \, - \, \kappa \sum_{\vec{n}} \;
\delta_{x',x+\vec{n}} \; .
\nonumber 
}
in which the free tight-binding hopping contributions of the form  $e^{-\delta h} $ are linearized, in order to be able to work with sparse matrices, but the diagonal couplings to Hubbard field and chemical potential are not.

It is clear that Eq.~(\ref{eq:partfunc}) is sign-problem free at half filling, i.e.~for $\mu =0$, as 
$\det ( M M^\dagger)= |\det{M}|^2$. 
This is no longer true for $\mu \neq 0$. 
By writing
\begin{align}
\calz =& \pathint{\phi} 
\, \big|\det M(\phi,\mu) \big|^2 \, \frac{\det{M} 
         (\phi,\mu)}{\det{M}(\phi,-\mu)} \notag\\ 
&\times \exp \big\{-\frac{\delta}{2U} \sum_{t=0}^{N_t-1} \sum_{x,y}
\phi_{x,t}^2 \big\}~, \label{eq:partfunc4a}
\end{align}
we can consider the complex ratio of determinants with unlike-sign chemical potentials as an observable in the 
``phase-quenched'' theory (defined by the modulus of the fermion determinant) with partition function $\calz_\mathrm{pq}$ and obtain
\begin{equation}
  \frac{\calz(\mu)}{\calz_\mathrm{pq}(\mu)} \, = \, \Big\langle
  \frac{\det{M}(\phi,\mu)}{\det{M}(\phi,-\mu)}
  \Big\rangle_\mathrm{pq} \, . \label{eq:reweight}
\end{equation}
This ratio is unity for $\mu\to 0$ and is a measure of the severity
of the sign problem, as it quantifies the effectiveness of brute-force reweighting.

\begin{figure}  
\begin{center}
\includegraphics[width=0.97\linewidth]{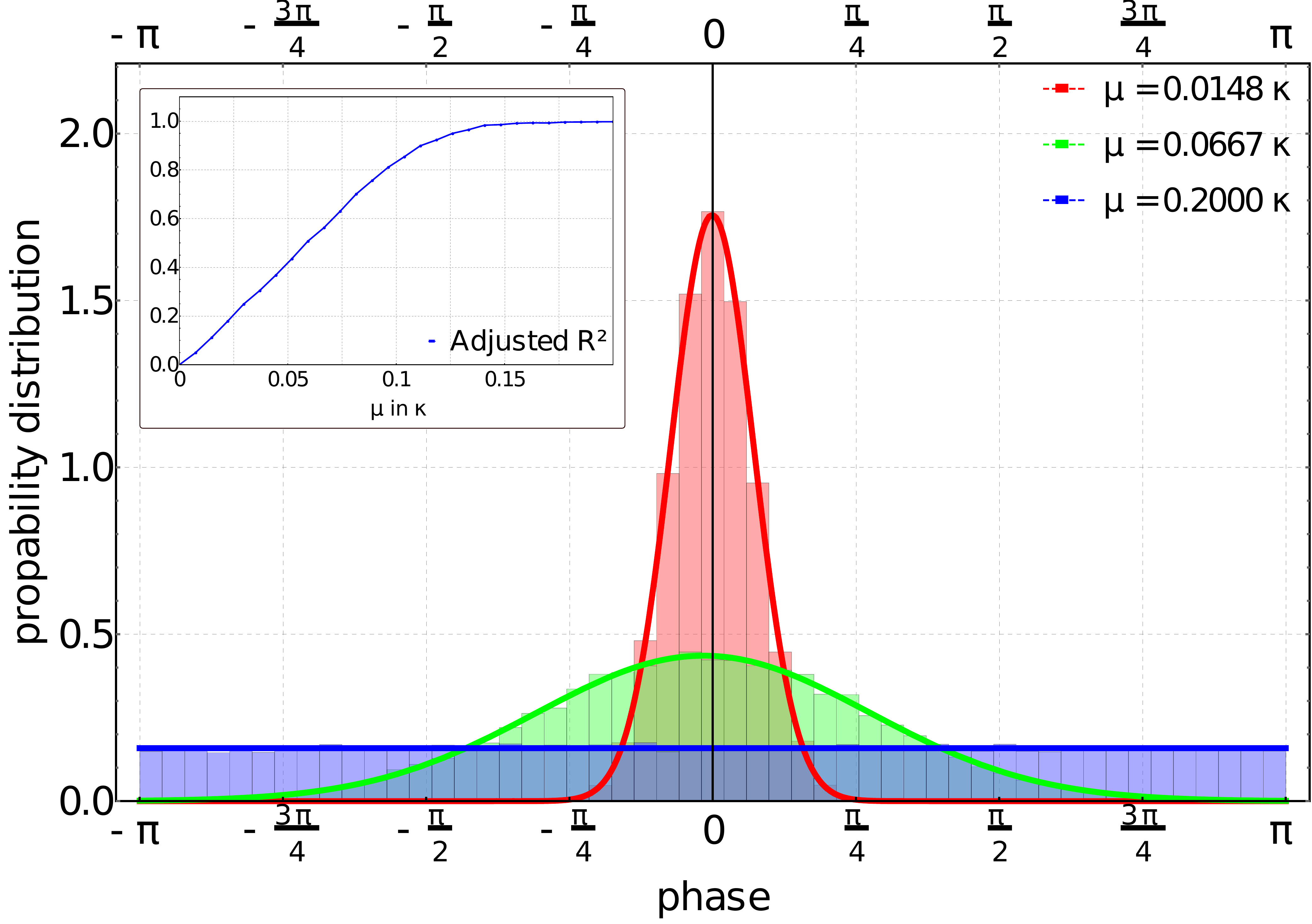}
\caption{Histograms of the phase of
  $\det{M}(\phi,\mu)/\det{M}(\phi,-\mu)$ for $N_s=N_t=6$, $\beta = 2.7 \kappa^{-1}$,
 $U=\kappa/2$ at  different $\mu$. The results are modelled with Gaussian (for $\mu = 0.0148\kappa$ and $0.0667\kappa$) and uniform
  (for $\mu = 0.2\kappa$) distributions respectively. The
  inlay shows the adjusted $\textrm{R}^2$ for a constant fit to
  data at different $\mu$. For $\mu \gtrsim 0.15\kappa$ 
the numerical data is well described by a uniform
  distribution, indicating a hard sign problem. An analogous figure as indication of a sign problem is obtained for graphene with long-range interactions \cite{Korner:2017qhf}. } 
\label{fig:signproblem1}
\end{center}
\end{figure}  

Fig.~\ref{fig:signproblem1} shows histograms of the 
phase\footnote{The modulus also deviates from unity at $\mu\neq 0$ but need not be considered here.} 
of the ratio $\det{M}(\phi,\mu)/\det{M}(\phi,-\mu)$ for different non-zero values of $\mu$, obtained on a lattice of $N_s\times N_s$ unit cells, with 2 sites per unit cell, and $N_s=N_t=6$, at $\beta = 2.7 \kappa^{-1}$ and $U=\kappa/2$, together with fit-model curves. The adjusted $R^2$ of a constant fit (corresponding to 
a uniform distribution) shows a rather rapid crossover and approaches values close to $1$ at $\mu
\approx 0.15\kappa$. This indicates that the signal is lost in
the noise rather quickly already on small lattices (signalling a hard sign problem), and for rather high 
temperatures and weak interaction strengths. This effect is enhanced with larger lattice sizes, lower temperatures 
and larger couplings. To present a quantitative comparison of brute-force
reweighting and the LLR method for different sysem sizes and interaction strengths is one of the main objectives
of this work. 

\section{Generalized density of the Hubbard field}\label{sec:doshubbard}

Assume we have a quantum system with
action $\beta S(\phi)$. Defining the density of states $\rho(E)$ as
\eqa{\rho(E)=&\pathint{\phi} \delta( S(\phi) - E )~,\label{eq:rhoedef}}
we can then express the partition function as
\eq{\calz = \int dE\,\rho(E) e^{\beta E}~,\label{eq:Zedef}}
and the vacuum expectation value of an observable $O(E)$ becomes 
\eq{\langle O \rangle = \frac{1}{\calz} \int dE\, O(E) \rho(E) e^{\beta E}~.\label{eq:vevdef}}
If $\rho(E)$ is known, then $\langle O \rangle$ can be obtained by (numerically or analytically)
integrating Eq.~(\ref{eq:vevdef}). Here we have assumed that we know how to express $O$ in terms of $E$, 
which is in general not the case however. 
Moreover, Eq.~(\ref{eq:rhoedef}) is ill-defined if $S(\phi)$
is not strictly real. To compute different observables in a generic setting, the concept of the density 
of states can thus be generalized to quantities other than the action. 

The basic idea of LLR is to obtain $\rho(X)$ (where $X$ is some observable) by carrying out a sequence of 
``microcanonical'' Monte-Carlo simulations, in which $X$ is forced to assume a set of different (sufficiently
dense) values $X_i$.
Obtaining the partition function or thermodynamic expectation values then essentially amounts to 
computing a Fourier or Laplace transform of $\rho(X)$. 
To alleviate the sign problem, $\rho(X$) must reflect the phase fluctuations 
of the path-integral measure. To this end, we consider $\rho(s=\Phi)$ in this work, 
where $\Phi$ is the spacetime-volume average of the auxiliary field, see below. 
First, we apply the transformation
\eq{ \phi_{x,t} \to \phi_{x,t} - i\mu~ \label{eq:mushift}}
to Eq.~(\ref{eq:partfunc}), which then leads to
\begin{align}
\calz(&\mu)=\nonumber\\
=&\pathint{\phi} \det{[M(\phi,\mu) M^{\dagger}(\phi,-\mu)]} \, 
\expo{ -\frac{\delta}{2U} \sum_{x,t} \phi_{x,t}^2} \nonumber\\
=&\pathint{\phi} \left|\det{ M(\phi,0)}\right|^2 \, 
\expo{ -\frac{\delta}{2U} \sum_{x,t} (\phi_{x,t}-i\mu )^2}. \label{eq:zmushift}
\end{align}
In this formulation, the complex part of the action has been shifted completely to the
bosonic sector. Eq.~(\ref{eq:zmushift}) is now rewritten as
\eqa{\calz(&\mu) =\pathint{\phi}  |\det{ M(\phi,0)}|^2 \, \nonumber\\ 
&\times \text{exp}\left\{ -\frac{\delta}{2U}  \sum_{x,t} \left( \phi_{x,t} - \Phi \right)^2  -  \frac{\delta V }{2U} \left( \Phi - i\mu  \right)^2 \right\}\,,}
where we have introduced the average Hubbard field
\eq{\Phi=\frac{1}{V} \sum_{x,t}  \phi_{x,t}~, \quad V= 2 N_c N_t~,\label{eq:meanphi}}
where $N_c$ denotes the number of unit cells with 2 sites each. Finally, we introduce 
\eqa{\rho(s)=&\pathint{\phi}  |\det{M(\phi,0)}|^2 \, \delta( \Phi - s ) \nonumber\\
&\times  \text{exp}\left\{ -\frac{\delta}{2U}  \sum_{x,t} \left( \phi_{x,t} - s \right)^2  \right\}~,\label{eq:rhosdef}}
and rewrite the partition function as
\eqa{\calz(\mu) = \int ds \, \rho(s) \,
\text{exp}\left\{ -  \frac{N_c}{UT} \left( s - i\mu  \right)^2 \right\} \, ,
\label{eq:zmu_rhos}
}
where $\rho(s)$ is the generalized density of states of the average Hubbard field $\Phi$ and will
be the target of our LLR calculation.

\section{Reconstructing the particle density}\label{sec:reconstruct}

Assume we have obtained $\rho(s)$ using some method.
Due to the oscillating contribution of the exponential, it is clear that Eq. (\ref{eq:zmu_rhos})
will be hard, if not impossible, to evaluate numerically. This is exacerbated by the fact
that LLR obtains $\rho(s)$ only for a discrete and finite set of points and with finite numerical 
precision. Our ultimate goal is to obtain the particle density $n(\mu)$. We present two reconstruction schemes which achieve
this in the following, which both operate in the in the frequency domain and avoid the instabilities of Eq. (\ref{eq:zmu_rhos}).

We note that $\rho(s)$ has a periodicity of $2\pi/\beta = 2\pi T $ and can thus be expanded in Fourier series. For later convenience we will first introduce  a dimensionless variable and density via
\begin{equation}
  x =\frac{s}{T} \, , \;\; \mbox{and} \quad \bar\rho(x) = T\rho(xT) \,. 
\end{equation}
If we furthermore introduce an imaginary chemical potential via
\begin{equation}
  \mu = i \theta T \, ,\quad \mbox{and} \;\; \calz^I(\theta) \equiv \calz(i\theta T)\, ,
\end{equation}
we observe that up to a Gaussian smearing with variance $U/2N_cT$ the generalized density of states is in fact essentially the same as the partition function at imaginary chemical potential,
 \begin{equation}
  \calz^I(\theta ) = \int dx\, \bar\rho(x) \, \text{exp}\left\{ -  \frac{N_cT}{U} \left( x - \theta  \right)^2 \right\} \, .
\end{equation}

We will obtain $\bar\rho(x)$ only at a discrete set of points $x_n=2\pi n/K$, where $n=\{0,\ldots,K-1\}$ and $\bar\rho_n \equiv \bar\rho(x_n)$. We must hence truncate the Fourier series, naturally leading to a discrete Fourier transform which can be used for interpolation via
\eqa{ \tilde{\rho}_k = \frac{1}{K} \sum_{n=0}^{K-1} \bar\rho_n \, e^{ 2 \pi i \, n k/K}   ~,\quad \bar\rho(x) \approx \sum_{k=0}^{K-1} \tilde{\rho}_k \, e^{- i k x} \, . \label{eq:fouriertrafo1} }
On the other hand, inserting (\ref{eq:fouriertrafo1}) into (\ref{eq:zmu_rhos}), we obtain
\eqa{\calz(\mu) & \approx \int dx\, \left(\sum_{k=0}^{K-1} \tilde{\rho}_k \, e^{- i k x} \right) \,
\text{exp}\left\{ -  \frac{N_cT }{U} \left(x - i\frac{\mu}{T}  \right)^2 \right\}.\nonumber\\
&= \sqrt{\frac{\pi U}{N_cT}} \, \sum_{k=0}^{K-1} \tilde{\rho}_k \, \exp\left\{ -  \frac{U}{4 N_c T}\,  k^2 -\frac{\mu}{T}\,  k   \right\}~, \label{eq:Zfugacity}
}
and the exact result is recovered for $K\to \infty$. In fact, in this limit, Eq.~(\ref{eq:Zfugacity}) becomes the fugacity expansion and we can identify for $k=N$,
\eqa{Z_c(T,N) =  \sqrt{\frac{\pi U}{N_cT}} \,   \tilde{\rho}_N  \, \exp \left\{ -  \frac{U}{4N_cT}\,  N^2 \right\}
\label{Zcan} }
as the corresponding  canonical partition function with particle number $N$. In the infinite volume limit $N_c \to \infty$ for fixed $N$, or equally so for $T\gg U$, we may therefore neglect the exponential factor and essentially identify the Fourier series coefficients $\tilde\rho_k $ of our generalized DOS with the canonical partition functions at $N=k$. At the same time, it is also evident from Eq.~(\ref{eq:Zfugacity}) that the generalized DOS itself then becomes equal, up to a constant factor, to the partition function at imaginary chemical potential,
 i.e. $\tilde\rho_k \propto Z_c(T,k)$, and $\bar\rho(\theta) \propto \calz^I(\theta)$ or   $\rho(s) \propto \calz(i s)$. 

 Moreover, one easily verifies that the truncated coefficients $\tilde\rho_k $ at finite $K$, obtained from the discrete Fourier transform in (\ref{eq:fouriertrafo1}), then yield pseudo-canonical partition functions, $\tilde\rho_k  \propto Z_c^K(T,k) $, which represent ensembles with particle number $N = k \mod K$. Likewise, the discrete sampling of $\bar\rho(\theta )$ provides us with an interpolation of $\calz^I(\theta)$ which agrees with the exact result for imaginary chemical potential at the discrete values $\theta_n = 2\pi n/K$.

 The general relation between $\rho(s)$ and the partition function at imaginary chemical potential of course
 also follows from Eq.~(\ref{eq:Zfugacity}), with $\mu = i s$ (and $K\to\infty $), 
\eqa{\calz(i s) &= \sqrt{\frac{\pi U}{N_c T}} \, \sum_{k=-\infty}^{\infty} \tilde{\rho}_k \, \exp \left\{ -  \frac{U}{4 N_cT} k^2 \right\} \, e^{ - i s k/T} \nonumber\\
  &\to  
  \sqrt{\frac{\pi UT}{N_c}} \; \rho(s) \, , \quad \frac{U}{N_c T} \to 0\,. }

In a finite volume, on the other hand, i.e.~at any finite number  $N_c$ of unit cells, the particle numbers $N$ are restricted to values between $\pm 2 N_c$, with $N=0$ at half filling, corresponding to an average of one of the maximally possible two electrons on each of the $2N_c$ sites. We then obtain the exact canonical partition functions from Eq.~(\ref{Zcan}) already for 
\[ K = K_\mathrm{max} = 4 N_c + 1 \, ,  \]
and with particle-hole symmetry at half filling, one actually only needs $K_\mathrm{max} = 2 N_c + 1 $.  

In principle, the particle number $N(\mu)$ can be directly obtained from Eq.~(\ref{eq:Zfugacity}), which is free of oscillating terms, by taking the deritvative with respect to $\mu$,
\eqa{N(\mu) &= T \frac{d}{d\mu} \ln \calz(\mu) \label{n_grandc}\\
  &= -\frac{\sum_{k=0}^{4N_c} \, k \, \tilde\rho_k \, \exp\left\{ -  \frac{U}{4 N_c T}\,  k^2 -\frac{\mu}{T}\,  k   \right\} }{\sum_{k=0}^{4N_c} \tilde\rho_k \, \exp\left\{ -  \frac{U}{4 N_c T}\,  k^2 -\frac{\mu}{T}\,  k   \right\} }\,.\nonumber }
Computing the $\tilde{\rho}_k$ from $\rho(s_n)$ can be done with high numerical precision using modern FFT libraries. 

Alternatively, we can also compute the chemical potential from the canonical partition functions, as the free energy difference of ensembles with subsequent particle numbers. From Eq.~(\ref{eq:Zfugacity}) we then obtain
\eqal{\mu(N+1/2) & =  -T  \big(\ln Z_c(T,N+1) -  \ln Z_c(T,N)\big)  \\ \approx & \; T \left[ \ln \tilde{\rho}_N -\ln \tilde{\rho}_{N+1} + \frac{U}{2N_cT} \big(N + 1/2 \big) \right]
~, \nonumber}{eq:mu_canonical}
and obtain the density in form of the number of particles per unit cell, $n(\mu) \equiv N(\mu)/N_c$, by inversion. The exact calculation would again require $K_\mathrm{max} = 2N_c +1$ Fourier coefficients.
This is then similar in spirit to Refs.~\cite{Alexandru:2005ix,deForcrand:2006ec,Nakamura:2015jra}, 
which carried out canonical calculations of QCD at finite charge density, 
or Ref.~\cite{PhysRevD.86.054507} which followed essentially the same strategy for finite isospin density from the lowest states in multi-pion correlators.
With truncating at $K < K_\mathrm{max}$, we strictly speaking obtain canonical ensembles at particle number $N$ modulo $K$ as discussed above. The  term $\propto U/2N_cT $ in Eq.~(\ref{eq:mu_canonical}) represents an explicit finite volume effect which, as we will discuss below, only contributes in trivial way and can be dropped. 

In tight-binding or mean-field calculations, there is no such term in the first place, and the generalized DOS can be calculated analytically. The result is of the form
\begin{equation}
  \ln\rho(s) = 2 N_c \int  d\varepsilon \, \rho_\varepsilon(\varepsilon) \, \ln\left(\cosh^2\frac{\varepsilon}{2T} - \sin^2\frac{s}{2T}\right)\, , \label{tblnrho}
\end{equation}
where $\varepsilon \ge 0$ is the single-particle energy with spectral density $\rho_\varepsilon(\varepsilon)$ for which an analytic expression is known in the infinite system \cite{PhysRev.89.662}. In a finite system with periodic (Born-von K\'arm\'an) boundary conditions we use the dispersion relation $\varepsilon = \varepsilon(\mathbf k)$ instead, and simply sum over the corresponding discrete set of points $\mathbf k_n$ within the first Brillouin zone, with energies $\varepsilon_n  = \varepsilon(\mathbf k_n) $. The same can be done to compute the exact density in the finite system with $N_c$ unit cells which then yields for the number of particles per unit cell,
\begin{equation}
  n(\mu) = \frac{1}{N_c} \sum_n \Big( \tanh\frac{\varepsilon_n+\mu}{2T} -  \tanh\frac{\varepsilon_n-\mu}{2T} \Big) \, .\label{tbdens}
\end{equation}
  
We have carried out a set of benchmark calculations in which we compared the 
canonical and grand-canonical reconstruction schemes. 
Thereby, a discete set of values $l_n = \ln\rho(s_n)$ for $s_n = 2\pi T n/K$, $N=\{0,\dots K-1\}$, was produced as mock data from the tight-binding calculation, which can efficiently be done with arbitrary numerical precision.
High-precision calculations are especially important in the reconstrucion of the density because we need with high precision the discrete Fourier transform of $\{\rho_n = e^{l_n}\}$ rather than that of $\{l_n\}$. The number density 
$n(\mu)$ was subsequently computed from the FFT result $\{\tilde \rho_k\}$, using both the fugacity expansion via (\ref{n_grandc}) and the canonical approach (\ref{eq:mu_canonical}). We have then compared both results with the exact calculation of the density based on the tight-binding formula (\ref{tbdens}). This was done for different setups, whereby the production of $\{l_n = \ln\rho(s_n)\}$ was done with different levels of floating point precision. The application of the reconstruction scheme was done with a $1024$ digit accuracy in each case to avoid additional errors. We find that both methods yield comparable results, with the canonical procedure having a very slight advantage for a given precision of $\ln\rho(s)$. We thus choose to use this procedure exclusively in
the following sections to process our LLR results. 

\begin{figure}[tb]
       \centering
\vspace{3mm}\includegraphics[width=0.40\textwidth]{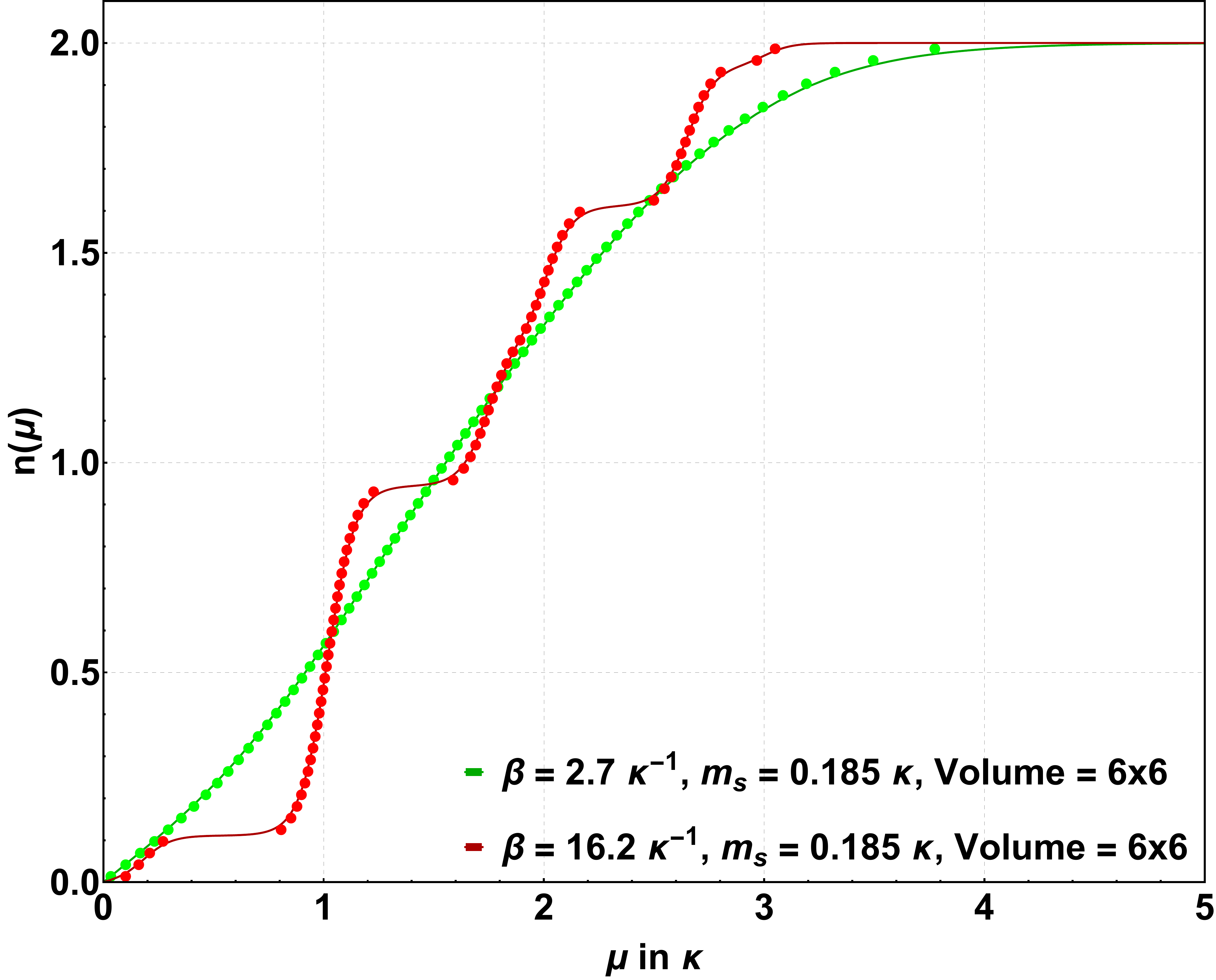}
        \caption{The number of particles $n(\mu)$ per unit cell on a lattice with $N_c=36$ unit cells for two different temperatures at $U=0$ from the tight-binding calculation (solid lines) and from the canonical reconstruction procedure based on (\ref{eq:mu_canonical}) using input data of $1024$ digit accuracy (discrete points).} 
\label{fig:mf_reconstruction}
\end{figure}

Fig. \ref{fig:mf_reconstruction} shows an example calculation of $n(\mu)$ for two different temperatures on a lattice with $N_c=36$ unit cells,  where $\ln\rho(s)$ was produced for $U=0$ with $1024$ digit accuracy and processed using (\ref{eq:mu_canonical}). This illustrates that our method can in principle
cover the entire width of the valence band, from the empty valence band at half fillg up to saturation when it is completely filled. The van Hove singularity will emerge at $\mu = \kappa$ in the infinite volume limit which can here be anticipated already by the rapid increase in the number density at the lower temperature around $\mu = \kappa$.

In practice, the leading source of errors is of course the precision with
which the Fourier coefficients $\tilde{\rho}_k$ can be obtained, which in turn is highly sensitive
to statistical errors of $\ln\rho(s)$ in our LLR calculations. In order to get to saturation, with a completely filled lattice, we would obviously need the maximum number $K_\mathrm{max} = 2N_c+1 $ of coefficients. So the double challenge here will be to compute as many of them as accurately as possible. 

\section{LLR results}\label{sec:llr_results}

\subsection{The algorithm}

The goal of LLR is to calculate derivatives of $\ln\rho(s)$ at
a sufficiently dense set of supporting points with high precision and to reconstruct $\rho(s)$ by integration. 
We divide the domain of 
support of $\rho(s)$ into $K$ intervals of size $\delta_s$. At the center of each of these intervals, the slope
$a_k=\frac{d}{ds}\ln\rho(s)|_{s=s_k}$ can be calculated from a stochastic non-linear equation \cite{Langfeld:2012ah}. 
A key element of this equation is the restricted and reweighted expectation 
value\footnote{The double-bracket notation is customary in the LLR 
literature and should be understood as defined by Eq.~(\ref{eq:llrev}). 
It is not implied here that an expectation value is taken twice.}
\begin{align}
\langle\langle W(\Phi)(a) \rangle\rangle_k =& \frac{1}{\calz_{LLR}} \int \mathcal{D}\phi\, \theta_{[s_k,\delta_s]}(\Phi) 
|\textrm{det} M(\phi)|^2 \nonumber\\
&\times W(\Phi) e^{-\beta S(\phi)} e^{-a\Phi}~. \label{eq:llrev}
\end{align}
Here $\calz_{LLR}$ is a normalization constant, $\Phi$ was introduced in Eq.~(\ref{eq:meanphi}), 
$a$ is an external parameter and $\theta_{[s_k,\delta_s]}$ is a window function which restricts $\Phi$ to an interval of 
size $\delta_s$ around $s_k$. 

With the choice $W(\Phi)=\Phi-s_k$, the coefficients $a_k$ are solutions of 
\eq{\langle\langle W(\Phi)(a_k) \rangle\rangle_k =0~.}
This equation can be solved through Robbins-Monro iteration 
\cite{MonroRobbins}: The sequence 
\eq{a^{(n+1)}_k=a^{(n)}_k +\frac{\alpha_n}{\delta^2_s}  \langle\langle W(\Phi)(a^{(n)}_k) \rangle\rangle_k \label{eq:monrob}}
converges to the correct result for any choice of $\alpha_n$ that fulfills
\eq{ \sum_{n=0}^\infty \alpha_n = \infty~,\quad \sum_{n=0}^\infty \alpha_n^2 < \infty~.}
This is true, even if $\langle\langle W(\Phi)(\cdot) \rangle\rangle_k$ is approximated by an estimator, as we do in Monte-Carlo calculations.
Moreover, if the iteration is terminated at some finite number $N$ and repeated many times, the final values $a_k^{(N)}$ are Gaussian
distributed around the true value $a_k$ and can be processed by a 
standard bootstrap analysis. 

The window function can be chosen in different ways. The straight-forward choice is a step function, but for HMC a Gaussian window function is
more appropriate, as its derivative can be taken, which implies that its
effect can be reproduced by a molecular-dynamics force term. In this work, we choose
\eqa{\langle\langle \Phi- & s\rangle\rangle(a) = \frac{1}{\calz_{LLR}}\pathint{\phi}   \det{M(\phi)} \det{M^{\dagger}(\phi)}\, ( \Phi - s) \, \nonumber \\ 
& \times \text{exp}\left\{ -\frac{\delta}{2U}  \sum_{x,t} \left( \phi_{x,t} - s \right)^2  -  \frac{ 1} {2 \delta_s^2  } 
\left( s - \Phi \right)^2    - a \Phi \right\}~,\label{eq:llr_action}}
where
\eqa{\calz_{LLR}&(a) = \pathint{\phi}  \det{M( \phi)} \det{M^{\dagger}( \phi)}\, \nonumber \\ 
& \times\text{exp}\left\{ -\frac{\delta}{2U}  \sum_{x,t} 
\left(  \phi_{x,t} - s \right)^2  -  \frac{ 1} {2  \delta_s^2  } \left( s - \Phi \right)^2    -  a \Phi \right\}.\label{eq:llr_partfun}}
The full procedure is then 
summarized as follows:
\begin{enumerate}
\item[1)] For a given $s_k$, initialize $a_k$ with some random value $a_{k}^{(0)}$ not too far from zero. 
\item[2)] 
Initialize Hubbard field (e.g with a value
which minimizes the window function). 
\item[3)]  With fixed $a_k$, thermalize Hubbard field with HMC trajectories according to Eq.~(\ref{eq:llr_partfun}), i.e. $\calz_{LLR}(a_k)$.
\item[4)]  With additional HMC trajectories, compute an estimate of $\langle\langle \Phi- s\rangle\rangle(a_k)$.
\item[5)]  Update $a_k$ using Eq.~(\ref{eq:monrob}).
\item[6)]  Continue from step $3$.
\end{enumerate}
In practice, we start with several repetitions of steps $3-6$ with fixed $\alpha=1$ and switch to \emph{underrelaxed} interations 
with $\alpha_{n+1}=\alpha_n /(1+n)$ after some time. Also, the whole procedure is terminated after some finite 
iteration number $N$ and repeated several times for each fixed $s_k$, to produce a sample of final $a_k^{(N)}$ values.

\begin{figure}[tb]
\vspace{3mm}\includegraphics[width=0.45\textwidth]{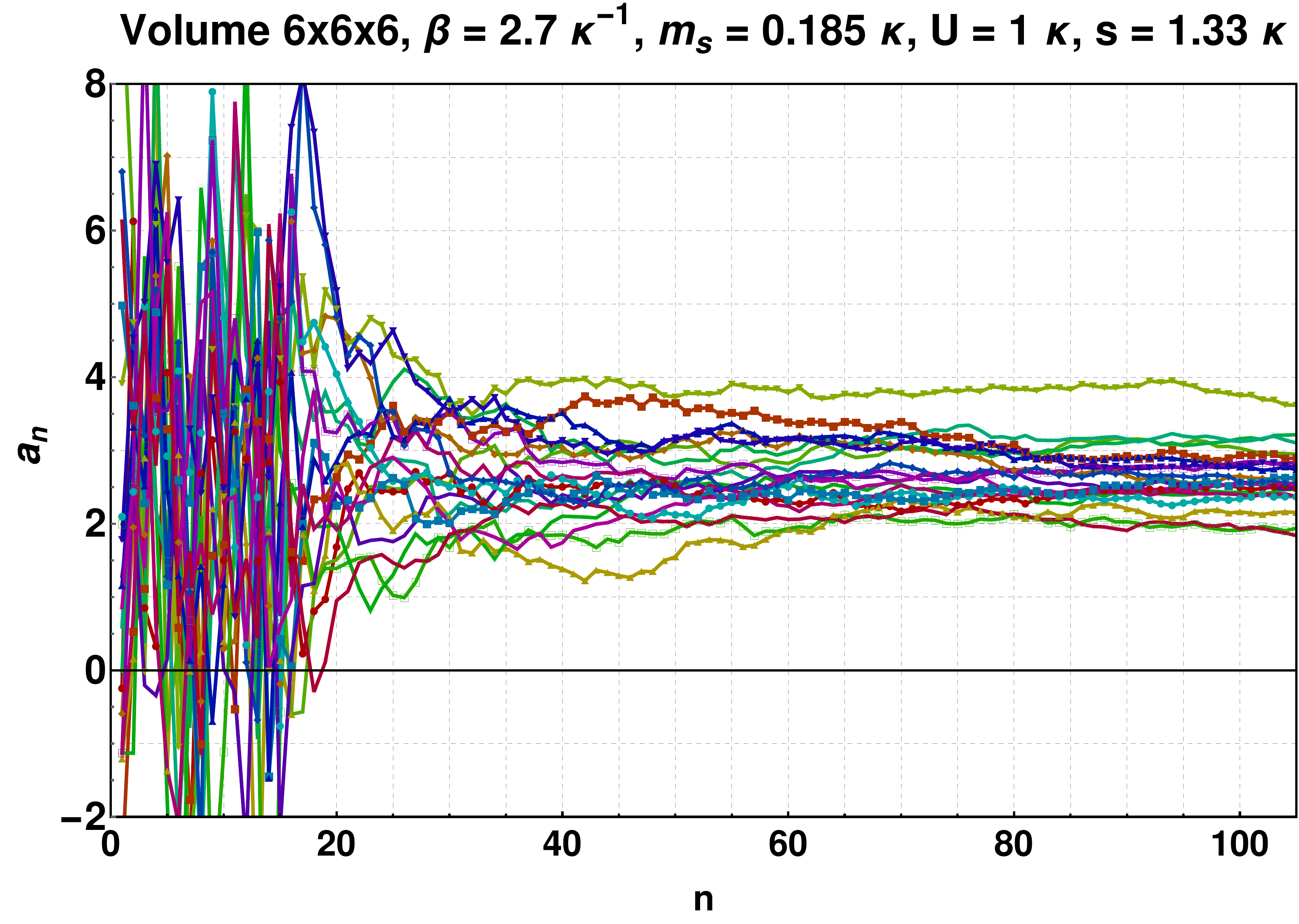}\\
       \caption{Illustration of stochastic Robbins-Monro iteration. A set of
20 starting values $a_k^{(0)}$ are 
generated and are each updated
according to Eq.~(\ref{eq:monrob}). Underrelaxation is switched on
at $n= 15$. The procedure is
terminated at $n= 105$ to obtain the final values used
for bootstrapping.} \label{fig:an_time_history}
\end{figure}

Fig.~\ref{fig:an_time_history} shows one example of a stochastic 
Robbins-Monro iteration, taken from our actual
production runs, where the procedure described above is
applied for a fixed set of external parameters. We choose 
$N_s=N_t=6$, $\beta=2.7 \, \kappa^{-1}$, $m_s=0.185 \, \kappa$, 
$U=1.0\, \kappa$, $s=1.33\, \kappa$ for illustration.\footnote{Note that the phenomenological value of the hopping parameter in the tight-binding model for graphene typically is $\kappa \approx 2.7$~eV, so this would correspond to a temperature of $T\approx 1$~eV in graphene.}   For each 
set of parameters considered in this work, we first obtain 
such a sample of $a_{k}$ values. We then obtain $\ln\rho(s_k)$,
and by extension $\rho(s_k)$, $\tilde\rho_k$ and $n(\mu)$ together with errorbars, by feeding bootstrap averages of  
the final $a_{k}^{(N)}$ into 
\eq{
\ln\rho(s_k)=\sum_{i=0}^{k-1}a_i \delta_s+\frac{1}{2}a_k \delta_s~, \label{eq:integrate_rho}
}
computing the Fourier transform of $\rho(s_k)$ and applying the canonical reconstruction
scheme described in Sec.~\ref{sec:reconstruct}.

\subsection{$N_t$ dependence}

\begin{figure}[htb]
\vspace{3mm}\includegraphics[width=0.45\textwidth]{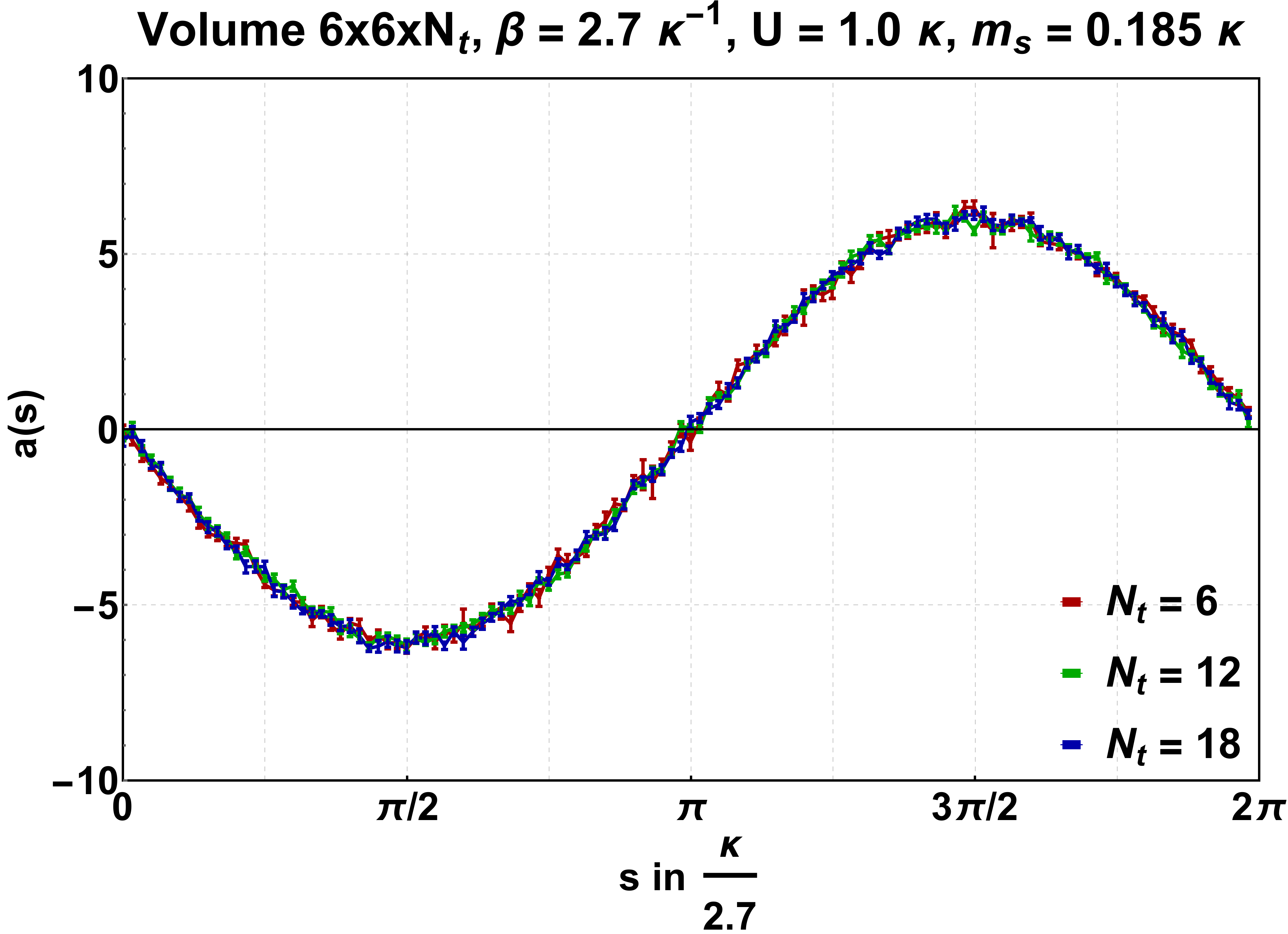}\\
\vspace{3mm}\includegraphics[width=0.45\textwidth]{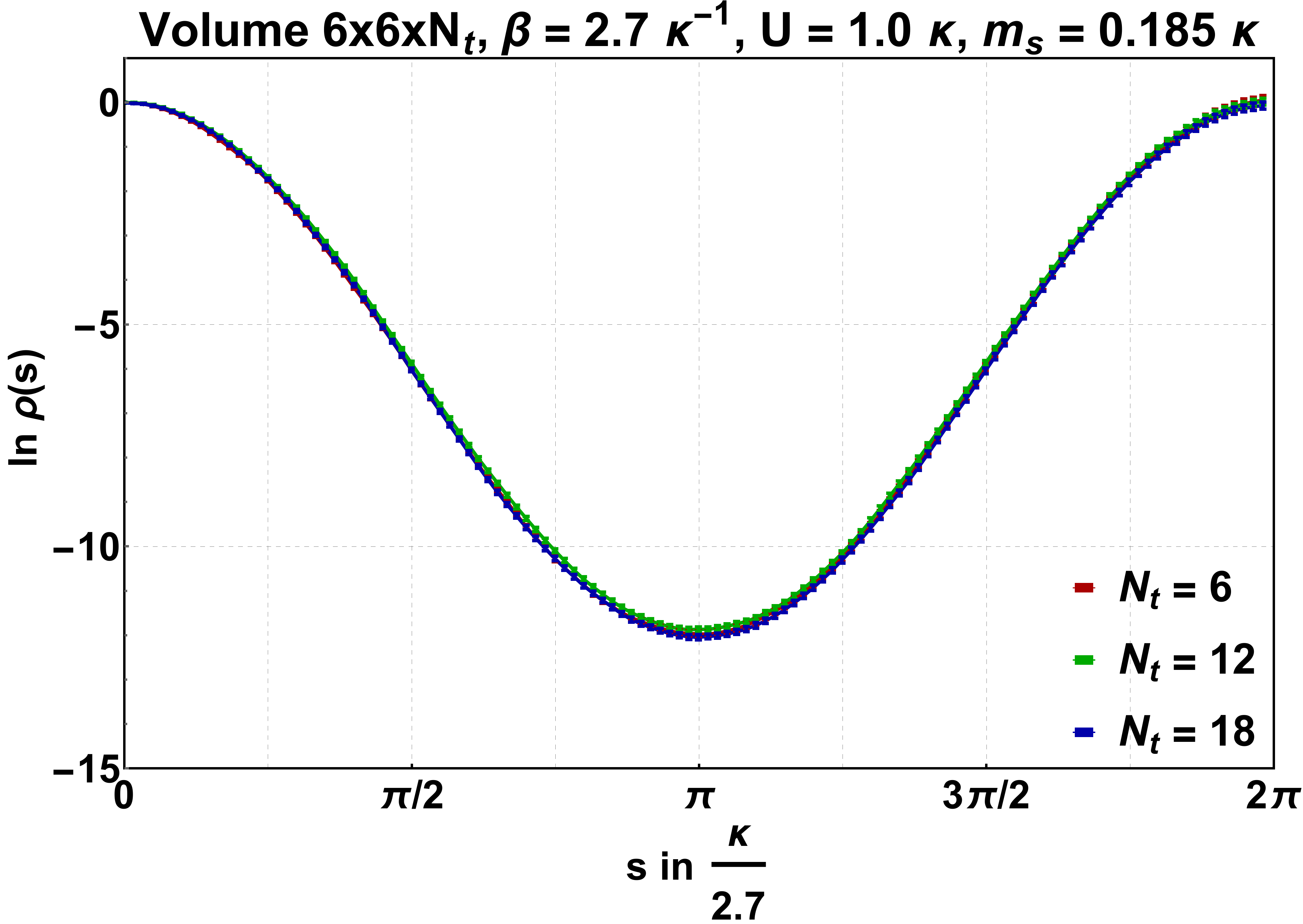}\\
\vspace{3mm}\includegraphics[width=0.45\textwidth]{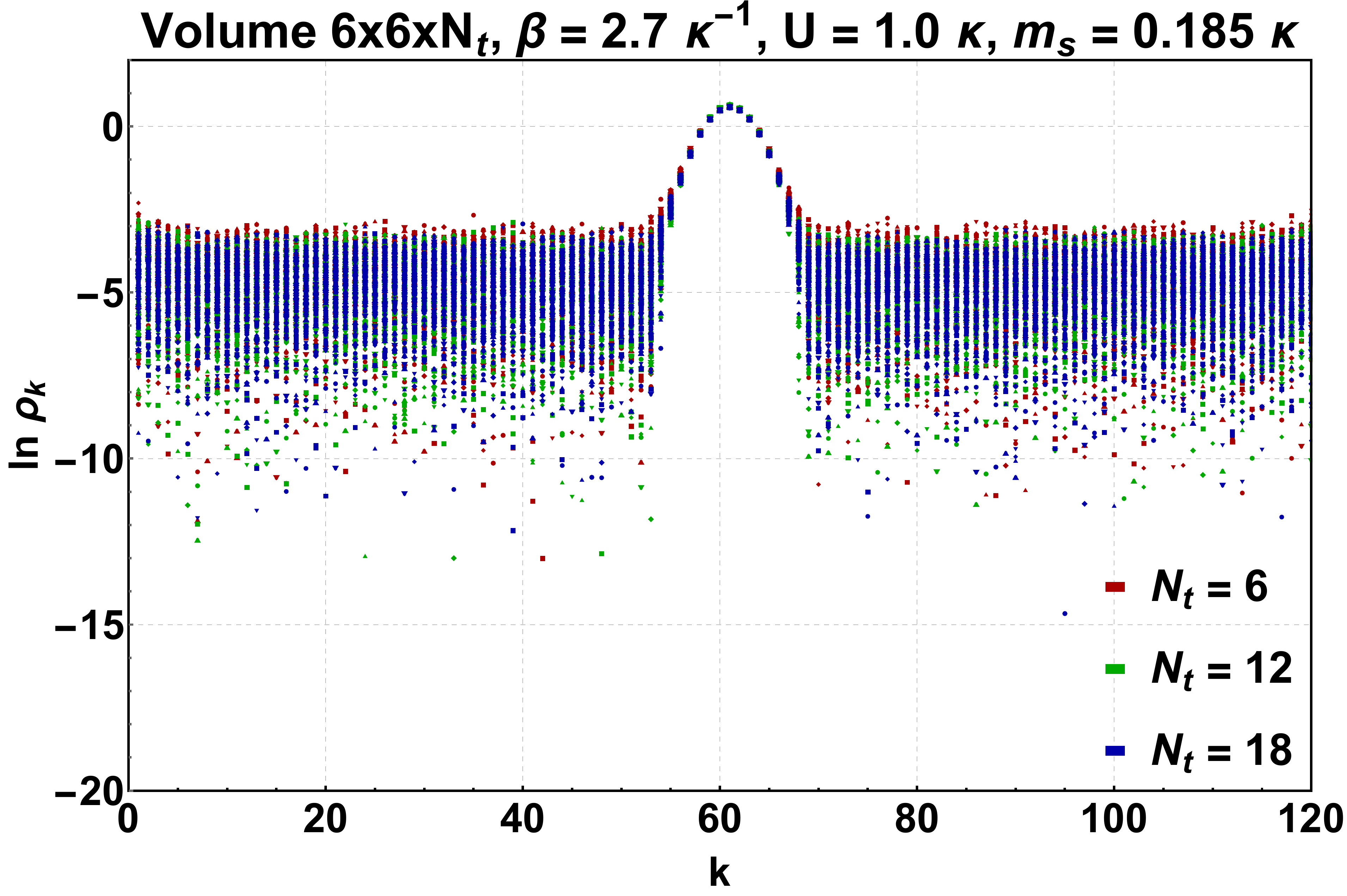}\\
       \caption{LLR result: $N_t$ dependence of $a(s)$, $\ln\rho(s)$, $\ln\tilde{\rho}_k$ for $N_s = 6$, 
$\beta = 2.7\, \kappa^{-1}$, $U=1.0\, \kappa$, $m_s= 0.185\, \kappa$. Individual bootstrap averages
are shown for $\ln\tilde{\rho}_k$ to illustrate loss of signal for the higher modes.} \label{fig:Nt_dependence}
\end{figure}

\begin{figure}[htb]
\vspace{3mm}\includegraphics[width=0.45\textwidth]{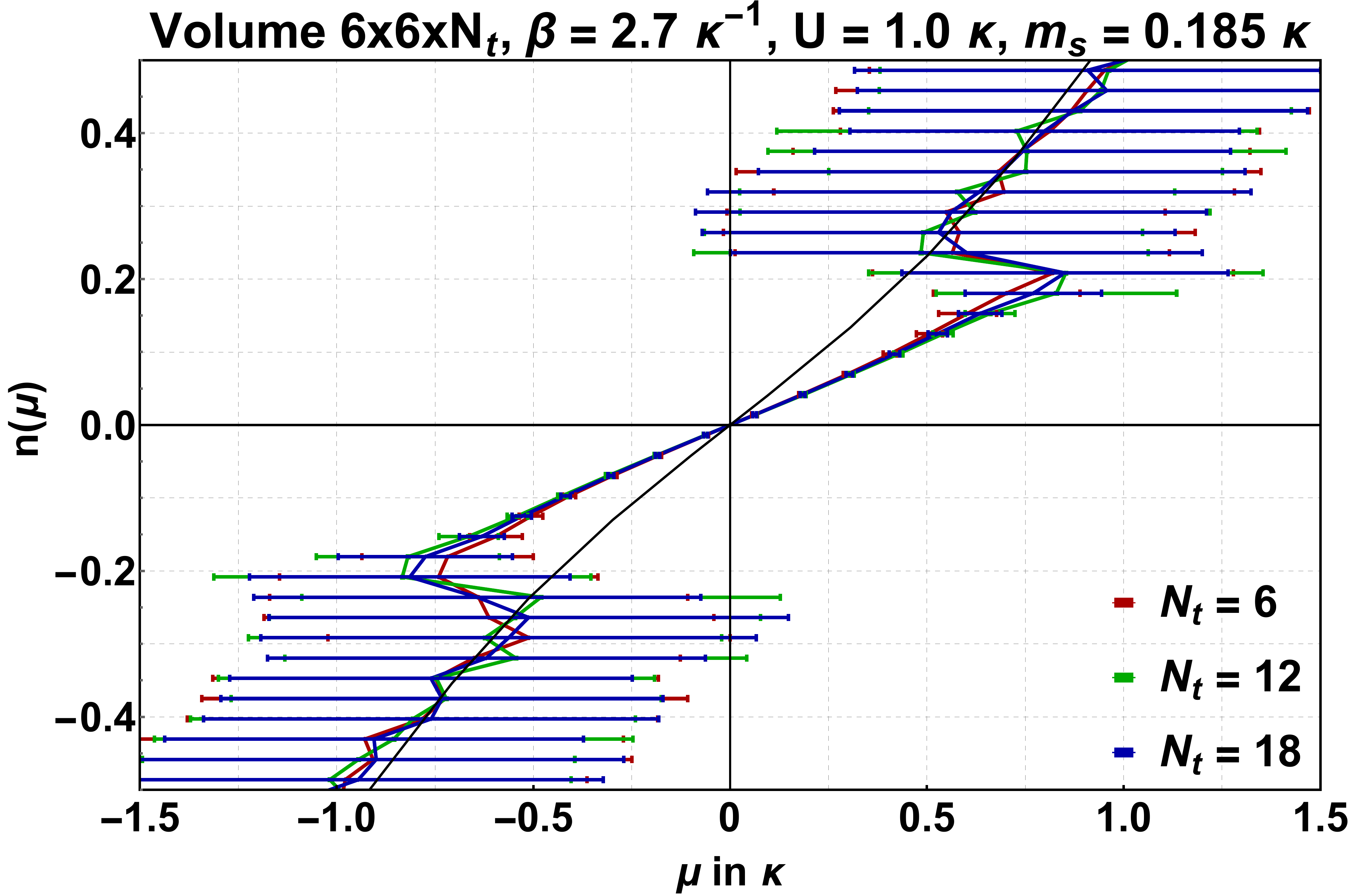}\\
\vspace{3mm}\includegraphics[width=0.45\textwidth]{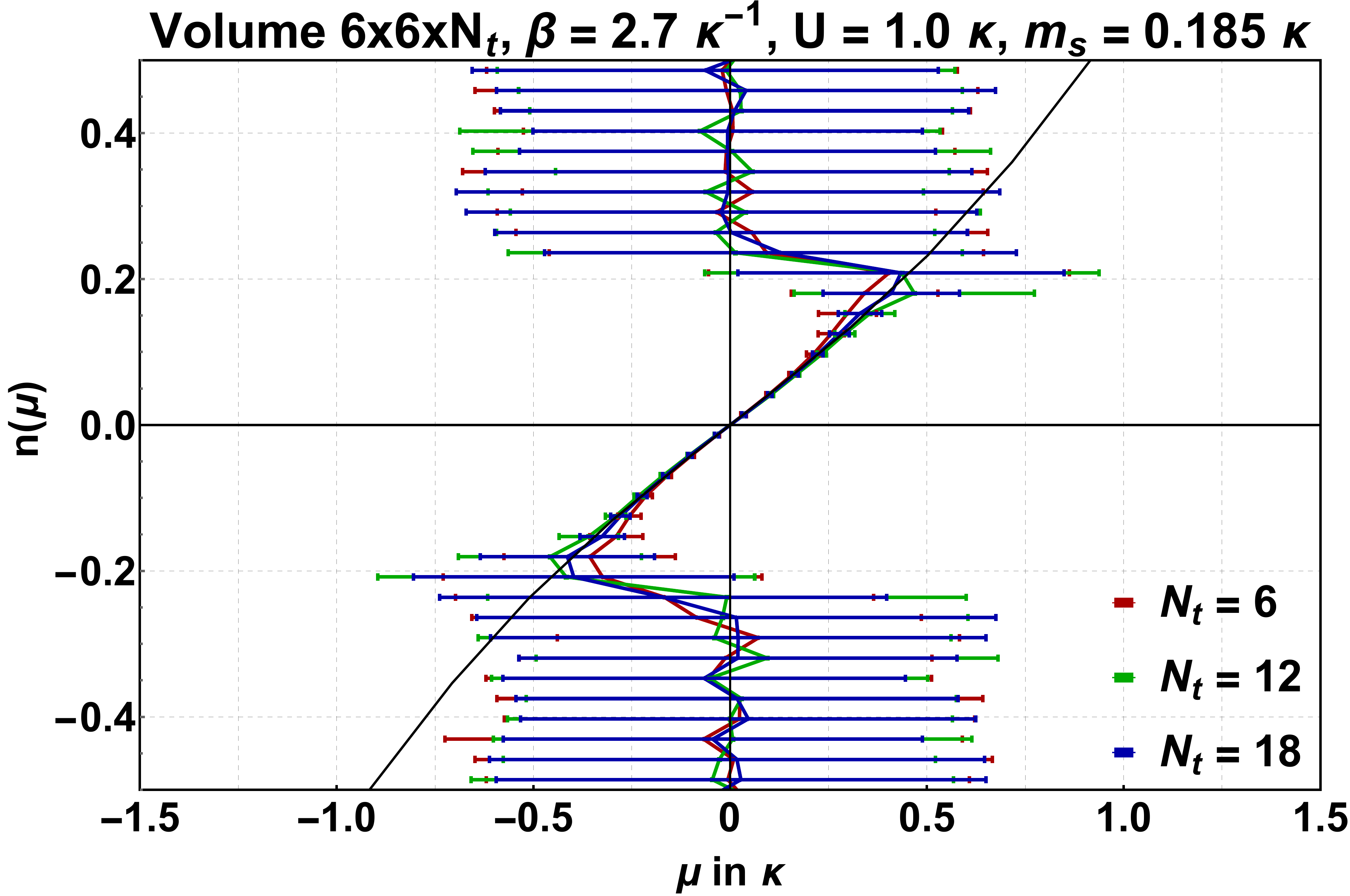}\\
       \caption{LLR result: $N_t$ dependence of $n(\mu)$ for $N_s = 6$, 
$\beta = 2.7 \,\kappa^{-1}$, $U=1.0 \,\kappa$, $m_s=0.185\, \kappa$.
Top figure includes the linear ($\sim U$) term in Eq.~(\ref{eq:mu_canonical}),
bottom figure does not.
Errorbars computed by boostrapping.
 Solid line shows the non-interacting tight-binding theory. 
 } 
\label{fig:Nt_dependence_density}
\end{figure}

We begin by studying the effect of the time-discretization $\delta$.
To this end, we carry out LLR calculations at $N_s = 6$, $\beta = 2.7\, \kappa^{-1}$, 
$U=1.0 \, \kappa$, $m_s=0.185\, \kappa$ for different values of $N_t$. 
Fig.~\ref{fig:Nt_dependence} shows the results for
$a(s)$, $\ln\rho(s_k)$ and $\ln\tilde\rho_k$, while the
final results for $n(\mu)$ are shown in Fig.~\ref{fig:Nt_dependence_density}. 
The latter figure includes two subfigures, whereby the linear $(\sim U)$
contribution to Eq.~(\ref{eq:mu_canonical}) is included or neglected respectively. 
Fig.~\ref{fig:Nt_dependence_density} also shows a corresponding calculation
of $n(\mu)$ in the non-interacting tight-binding theory. All errorbars were obtained
through bootstrap analysis. 

Our first observation is that the dependence on $N_t$ is very mild for our choice
of parameters. It is practically invisible in $a(s)$ and $n(\mu)$. A very small difference between
different $N_t$ can be seen in $\ln\rho(s_k)$ and $\ln\tilde\rho_k$, which is of a similar
magnitude as the statistical uncertainty however. On the other hand, 
our results clearly demonstrate exponential error suppression, whereby the relative error of 
$\ln \rho(s)$ is roughly the same across several orders of magnitude.  
We find that $\ln\tilde\rho_k$ is extremely sensitive to this small error
however, to a degree that only the first few Fourier modes $\ln\tilde\rho_k$ can be computed
accurately. This can be traced back to the fact that $\rho(s)$ enters into
the Fourier transform and not $\ln \rho(s)$. It is also reflected in our computation of $n(\mu)$, 
which exhibits a loss of signal at $\mu \approx 0.5 \, \kappa$, indicating the onset of a hard
sign problem. 

We note that for $U=1.0 \, \kappa$ which is well inside the weak-coupling phase of the model, and the temperature considered here, $n(\mu)$ basically fully agrees with the infinite-volume limit in the non-interacting theory when the linear term in Eq.~(\ref{eq:mu_canonical}) is dropped.  
We take this as an indication that this extra term represents the dominant finite-volume effect at finite $U$ which however is a rather trivial one to correct. 
Further confirmation of this is provided by a comparison between results from $N_s=6$ and $N_s=12$ 
lattices, which also reveals a faster convergence to the thermodynamic limit without this term. 
We thus drop this term for all results presented in the following. 
We expect that deviations from the non-interacting limit will become visible at stronger couplings, of course.
This is investigated further, and ultimately confirmed, below. 

\subsection{$m_s$ dependence}

\begin{figure}[tb]
\vspace{3mm}\includegraphics[width=0.45\textwidth]{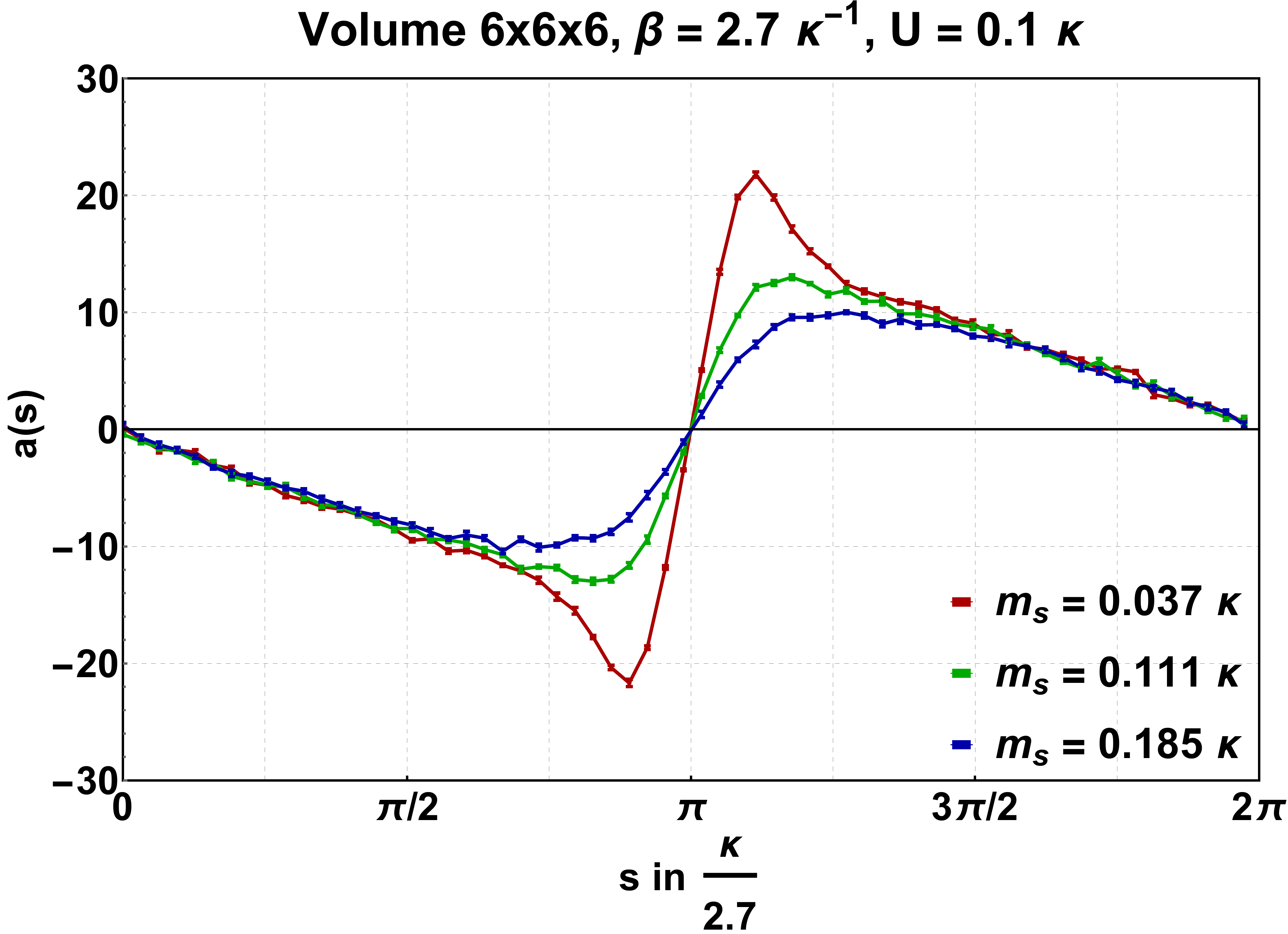}\\
\vspace{3mm}\includegraphics[width=0.45\textwidth]{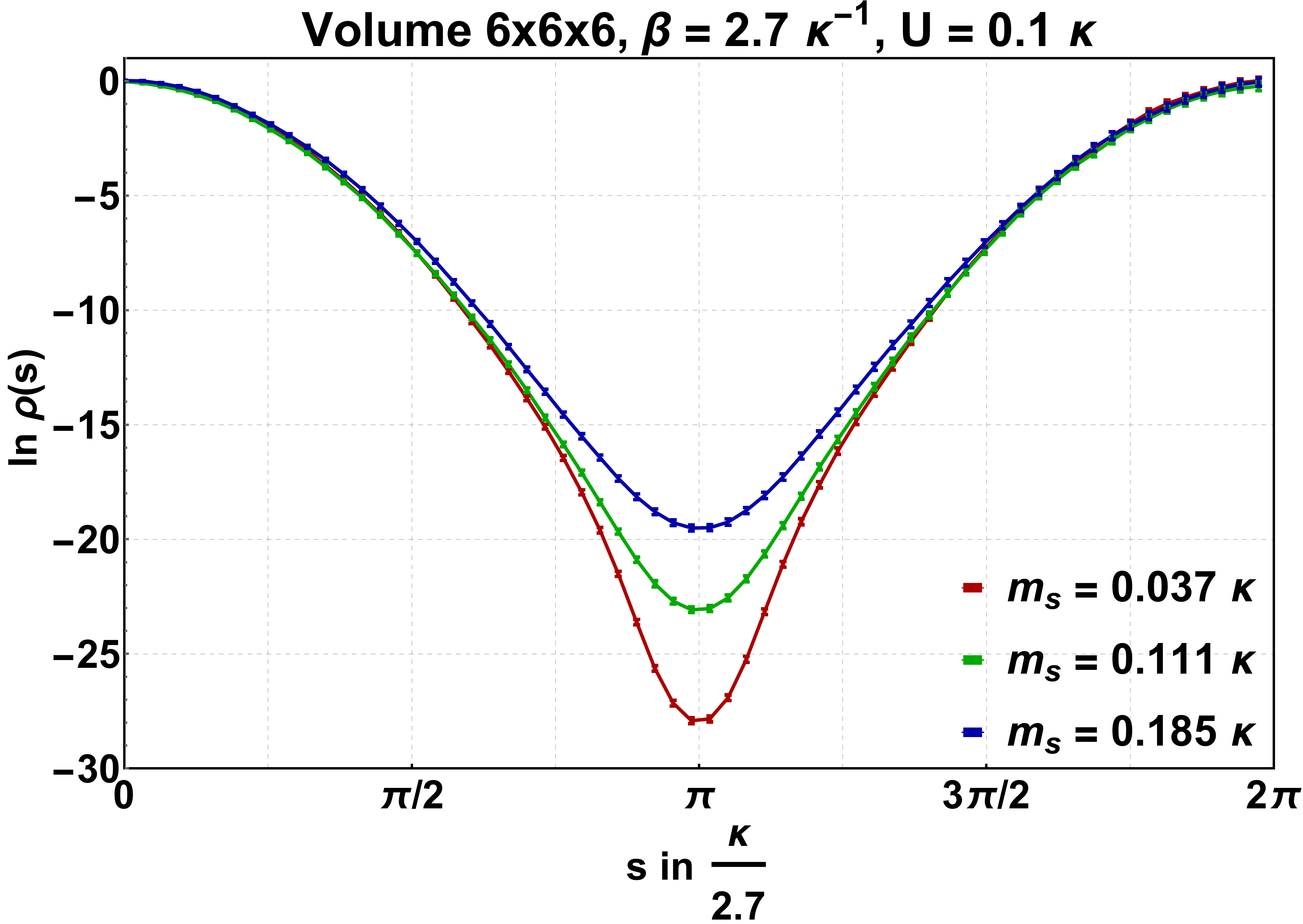}\\
       \caption{LLR result: $m_s$ dependence of $a(s)$ and $\ln\rho(s)$ for $N_s = N_t = 6$, 
$\beta = 2.7 \,\kappa^{-1}$, $U=0.1 \, \kappa$. } \label{fig:ms_dependence}
\end{figure}

We now turn to studying the dependence on the explicit sublattice and spin-staggered mass term $m_s$. Given that 
such a term already opens an explicit gap in the energy spectrum, we carry
out this study at the comparatively weak coupling strength of $U=0.1 \,\kappa$. 
We find that, again, the number density $n(\mu)$ coincides with the non-interacting
theory and shows no significant dependence on $m_s$. The linear term in Eq.~(\ref{eq:mu_canonical})
has a negligible effect here, due to the small 
value of $U$. Fig.~\ref{fig:ms_dependence} shows the results for
$a(s)$ and $\ln\rho(s)$ with 
$N_s = N_t = 6$, $\beta = 2.7 \, \kappa^{-1}$ and three different choices of $m_s$. 
We refrain from showing any additional figures for $\ln\tilde\rho_k$ and $n(\mu)$, 
as these fully agree (within statistical errors) with the results
shown in the lowest panels of Figs.~\ref{fig:Nt_dependence} and \ref{fig:Nt_dependence_density}.

An interesting observation here is that $m_s$ has a quite strong effect on both 
$a(s)$ and $\ln\rho(s)$, which turns out not to carry over to $n(\mu)$ at all. The underlying
reason is that this dependence is only present in regions where $\rho(s)$ is strongly
suppressed. It is only visible due to the logarithmic scale, and thus has no significant effect
on the computation of the Fourier modes.  

\subsection{$U$ dependence}

Having validated our numerical procedure at weak coupling, we now turn to a more detailed study of the dependence on the interaction strength $U$. This represents the central part of this work to which the bulk of our computing
resources were dedicated. We thereby computed $a(s)$, $\ln\rho(s)$, $\ln\tilde\rho_k$
and $n(\mu)$ again with $\beta = 2.7 \, \kappa^{-1}$, $m_s=0.185 \, \kappa$
for several different choices of $U$. To have control over finite volume and time-discretization effects
we have studied two different lattice sizes, $N_s = N_t = 6$ and $N_s = N_t = 12$, 
respectively. 

\begin{figure}[htb]
       \centering
\includegraphics[width=0.45\textwidth]{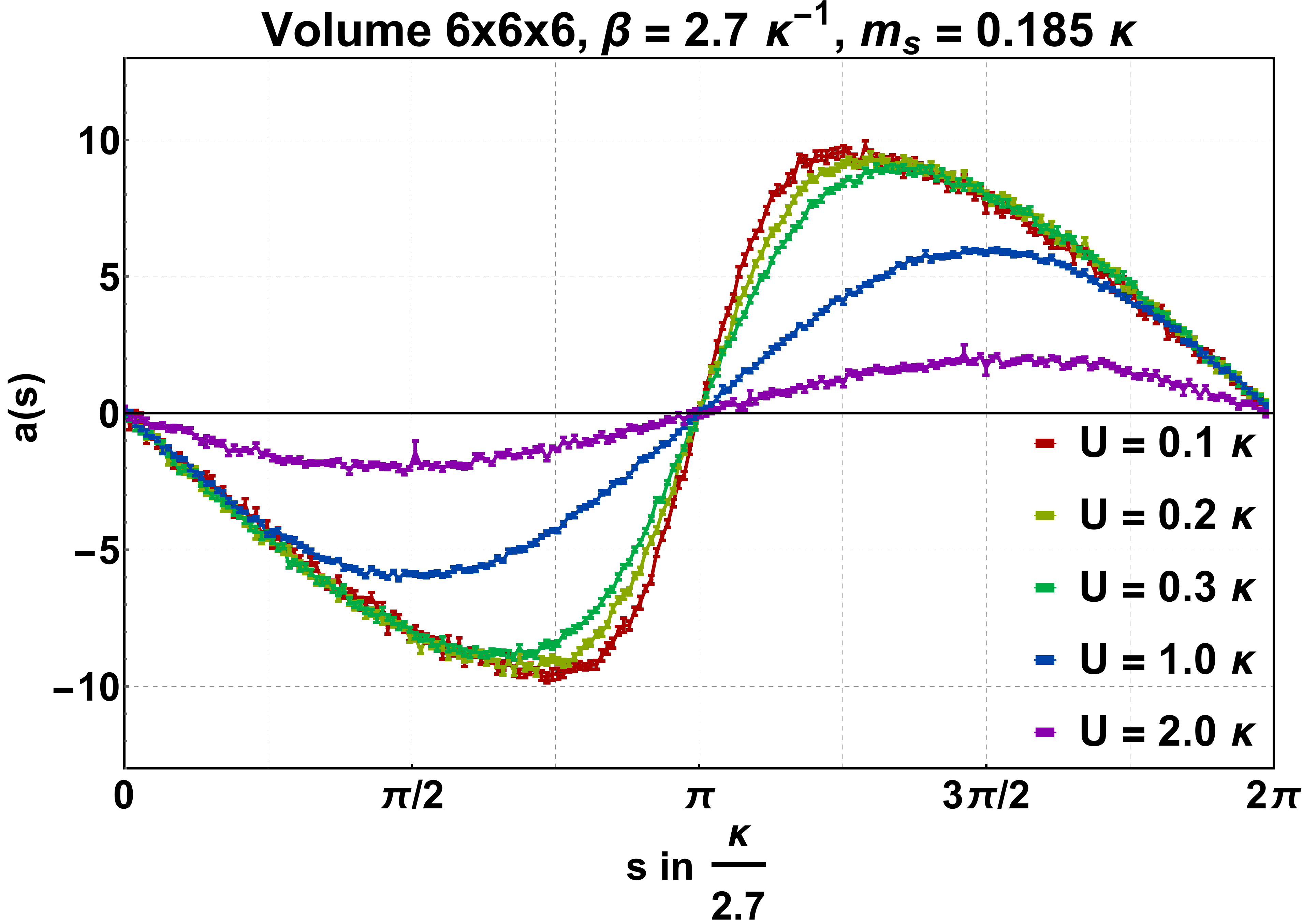}\\
\includegraphics[width=0.45\textwidth]{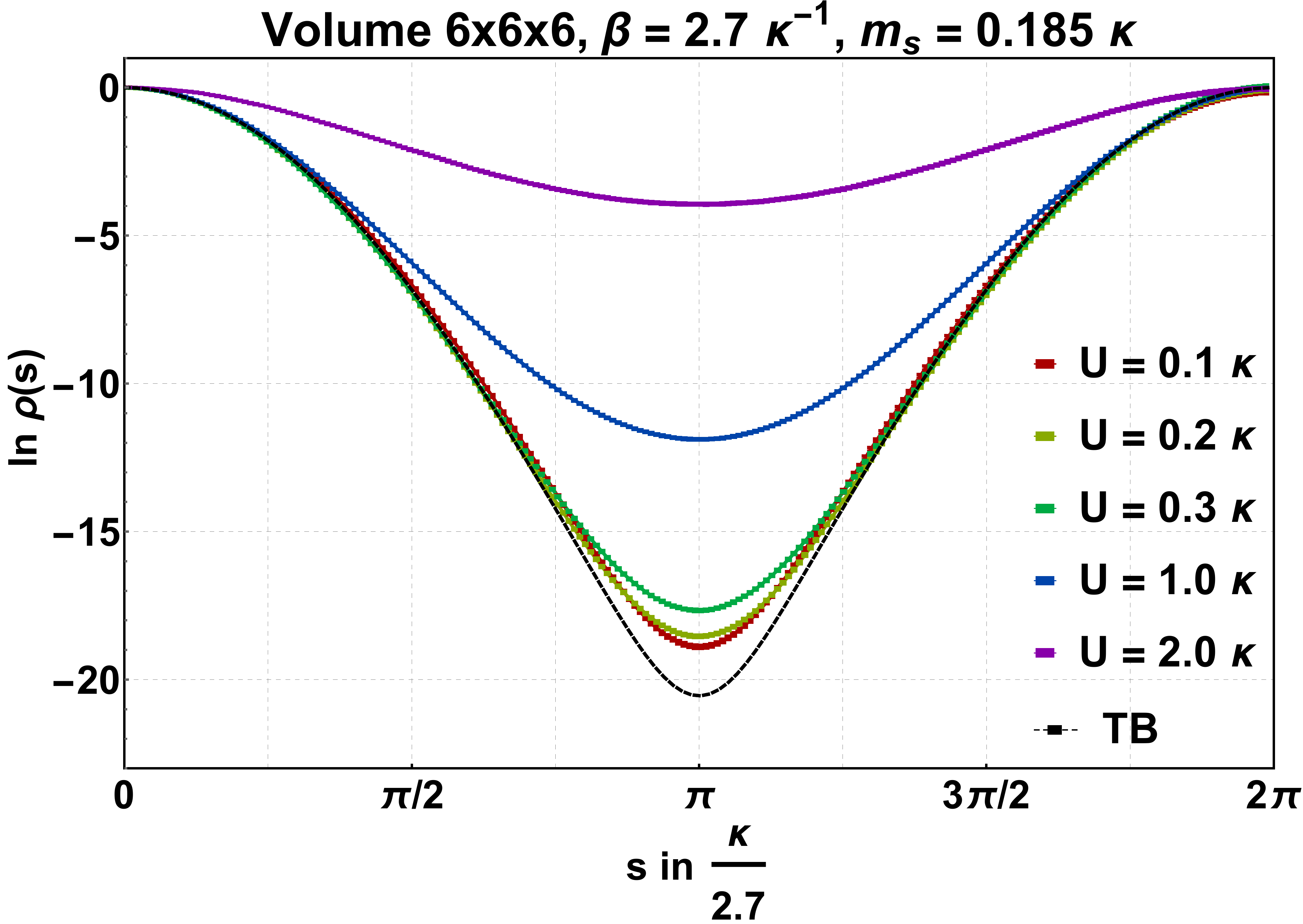}\\
 \includegraphics[width=0.45\textwidth]{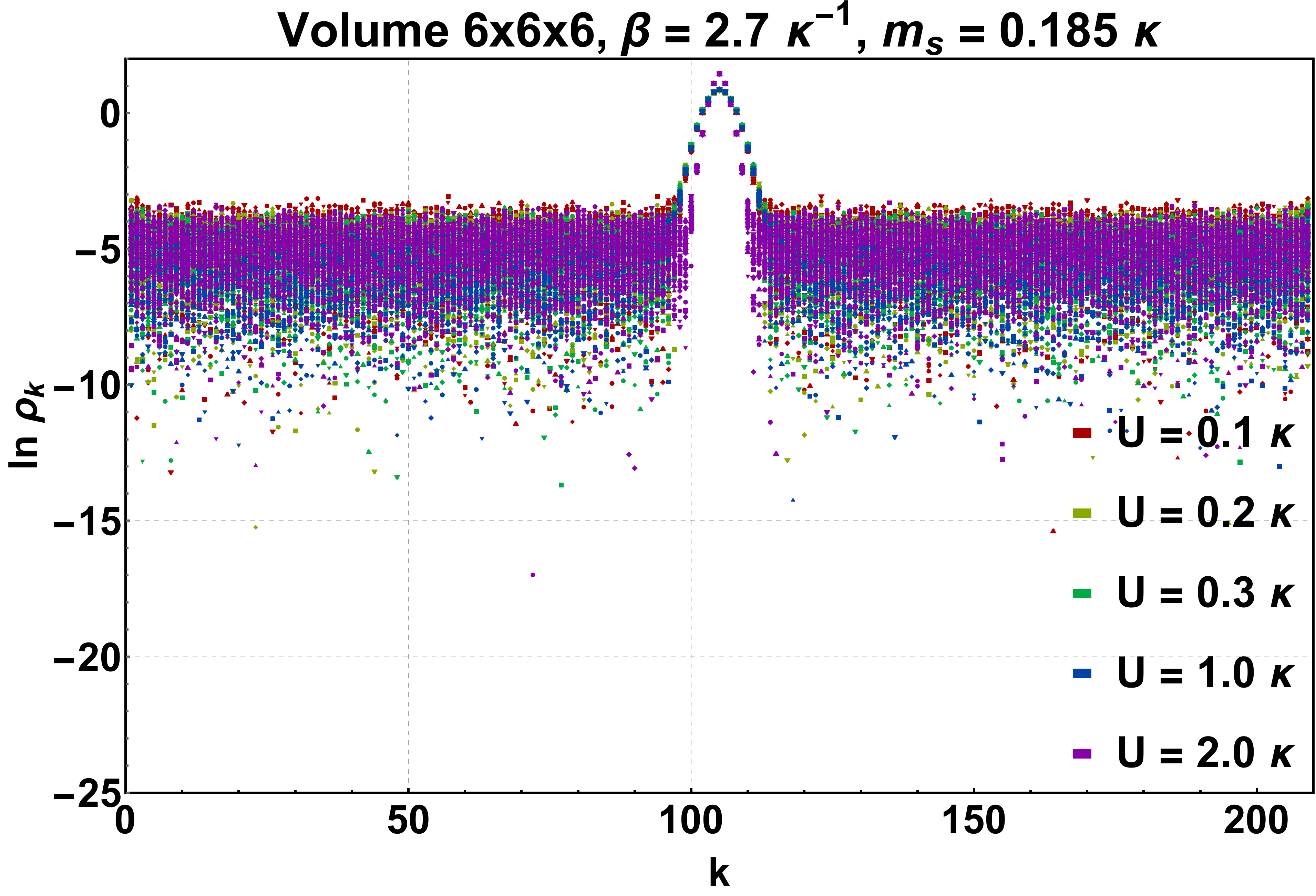}\\
       \caption{LLR result: $U$ dependence of $a(s)$, $\ln\rho(s)$, $\ln\tilde\rho_k$ for $N_s = N_t = 6$, 
$\beta = 2.7 \,\kappa^{-1}$, $m_s=0.185 \, \kappa$. Individual bootstrap averages
are shown for $\ln\tilde\rho_k$. Result for non-interacting tight-binding theory is included
for $\ln\rho(s)$.} \label{fig:U_dependence_vol6}
\end{figure}

\begin{figure}[htb]
       \centering
  \includegraphics[width=0.45\textwidth]{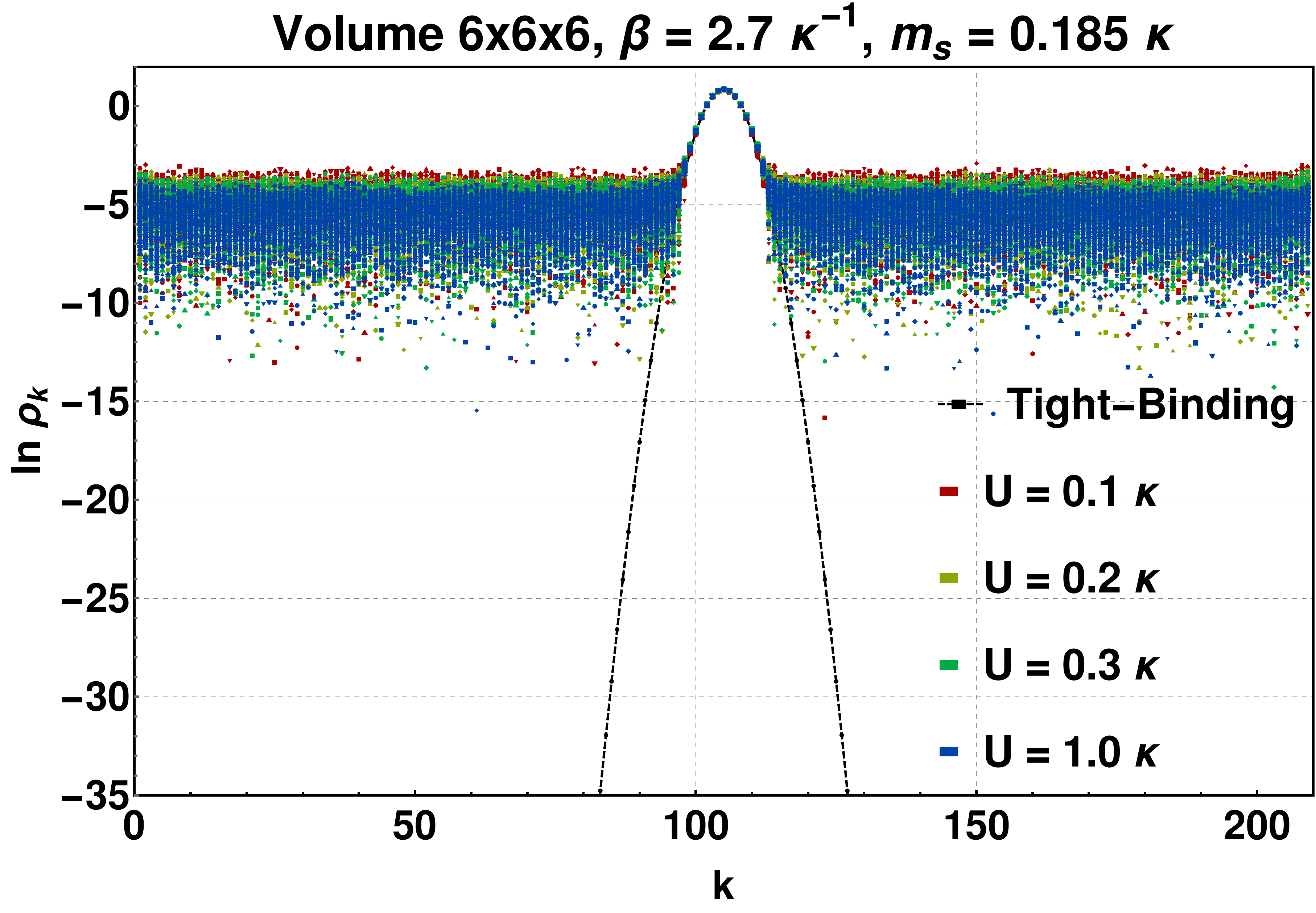}\\
        \caption{LLR result: $\ln\tilde\rho_k$ for $N_s = N_t = 6$, 
$\beta = 2.7\, \kappa^{-1}$, $m_s=0.185 \,\kappa$ and different $U$, compared with non-interacting tight-binding
theory. } \label{fig:U_dependence_vol6_tb}
\end{figure}

\begin{figure}[htb]
       \centering
  \includegraphics[width=0.45\textwidth]{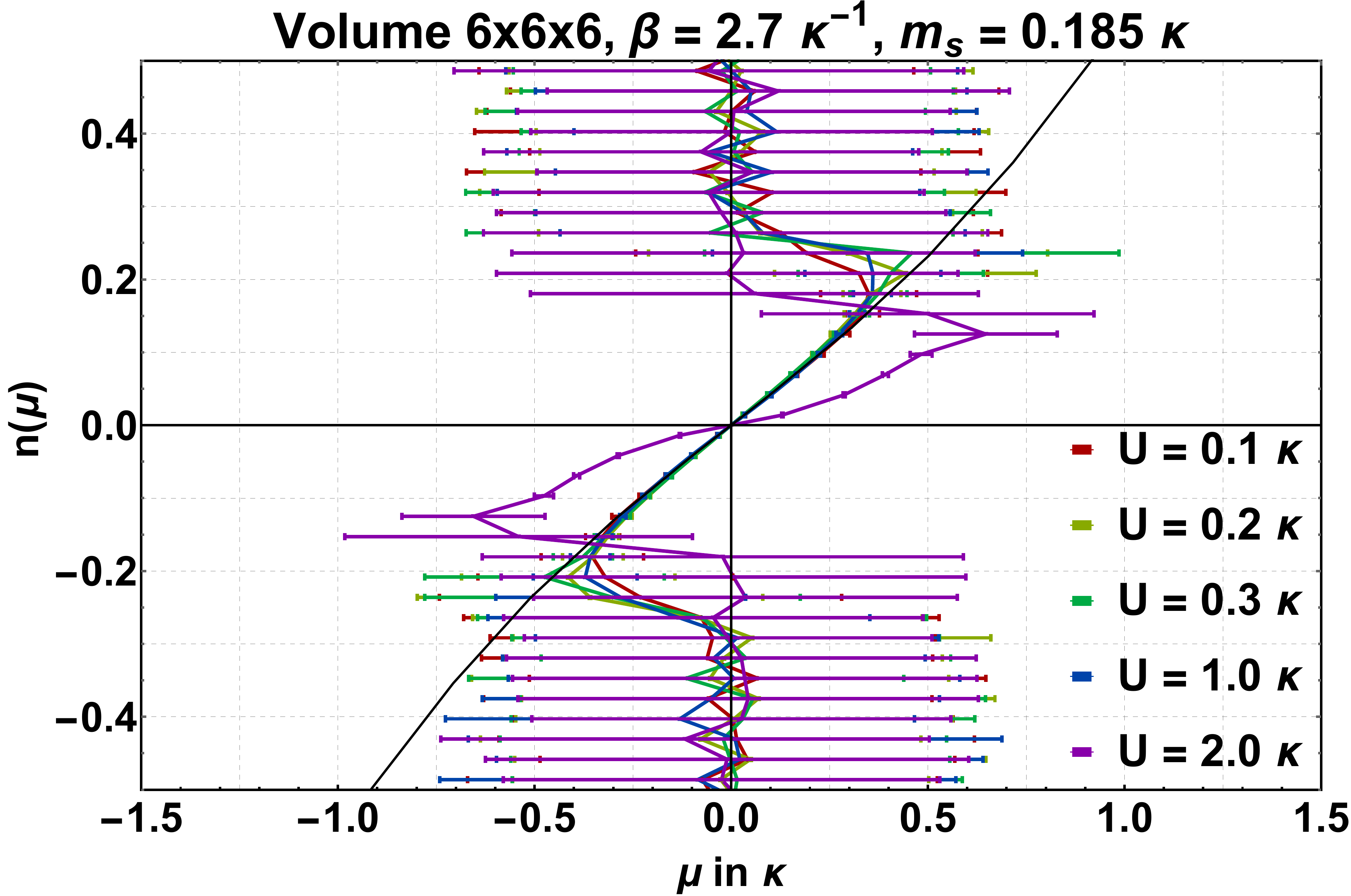}
       \caption{LLR result: $U$ dependence of $n(\mu)$ for $N_s = N_t = 6$, 
$\beta = 2.7\, \kappa^{-1}$, $m_s=0.185 \, \kappa$. Errorbars computed by boostrapping.
Solid line shows the non-interacting tight-binding theory. 
 } \label{fig:U_dependence_density_vol6}
\end{figure}

Fig.~\ref{fig:U_dependence_vol6} shows the $U$ dependence of $a(s)$, $\ln\rho(s)$ and $\ln\tilde\rho_k$ 
for $N_s = N_t = 6$, $\beta = 2.7 \, \kappa^{-1}$, $m_s=0.185\, \kappa$. 
For $\ln\rho(s)$ we include the tight-binding result to illustrate the approach to the non-interacting limit. 
The first observation is that $a(s)$ gets suppressed when $U$ is increased,
which ultimately makes simulations more expensive at strong coupling. On the other hand, we clearly see
a deviation from the non-interacing limit in the Fourier modes $\ln\tilde\rho_k$ for the strongest interaction
strength $U=2.0\, \kappa$. 
To underscore that this deviation is absent for all weaker interactions, we
show a seperate plot in Fig.~\ref{fig:U_dependence_vol6_tb} which directly compares $\ln\tilde\rho_k$ for $U\leq 1.0\, \kappa$ to the tight-binding theory. 
Our $N_s = N_t = 6$ results for $n(\mu)$ are shown in 
Fig.~\ref{fig:U_dependence_density_vol6}. They clearly show a corresponding drop of the number density at
fixed $\mu$ for the strongest coupling. 

\begin{figure}[htb]
       \centering
\includegraphics[width=0.45\textwidth]{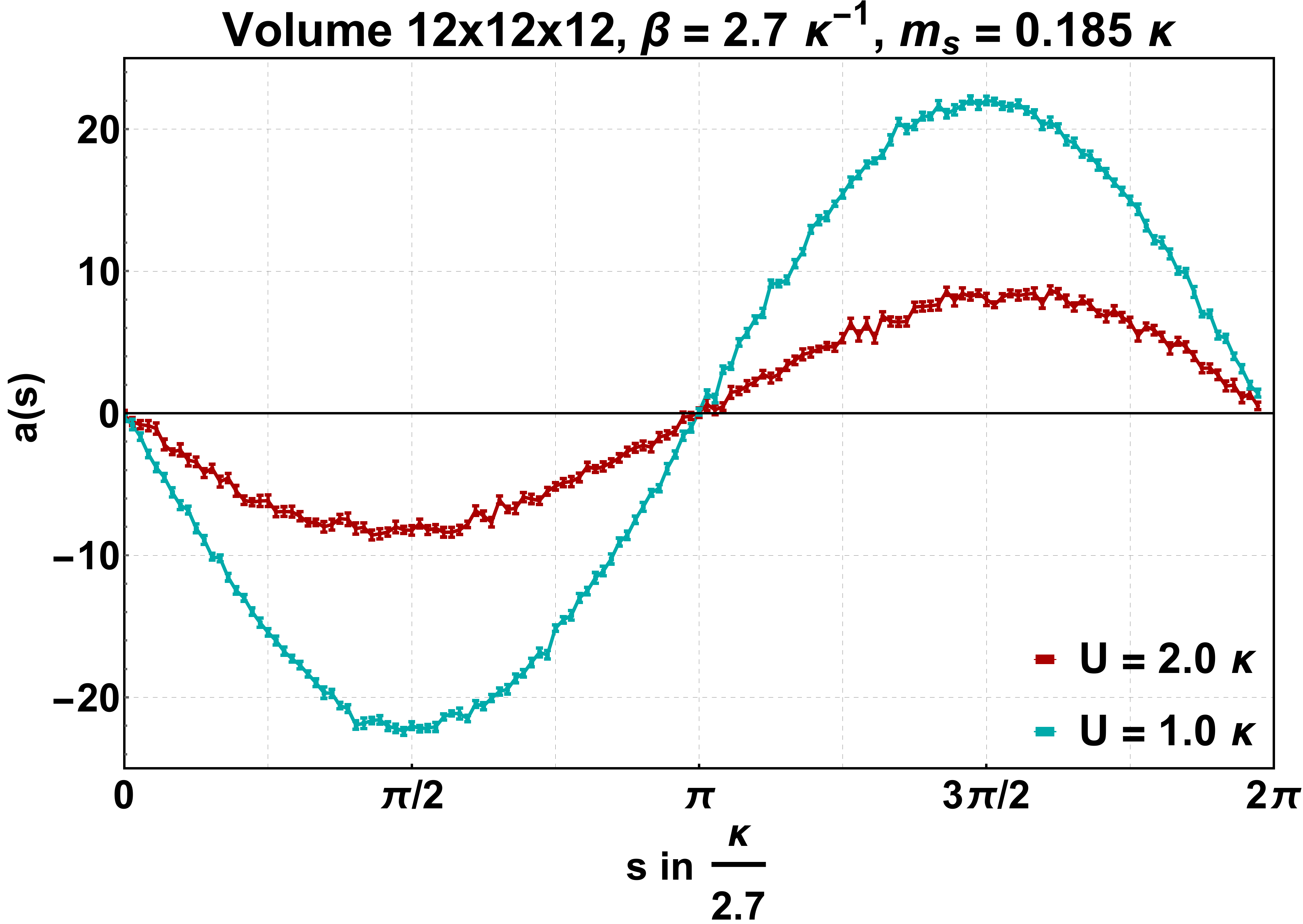}\\
\includegraphics[width=0.45\textwidth]{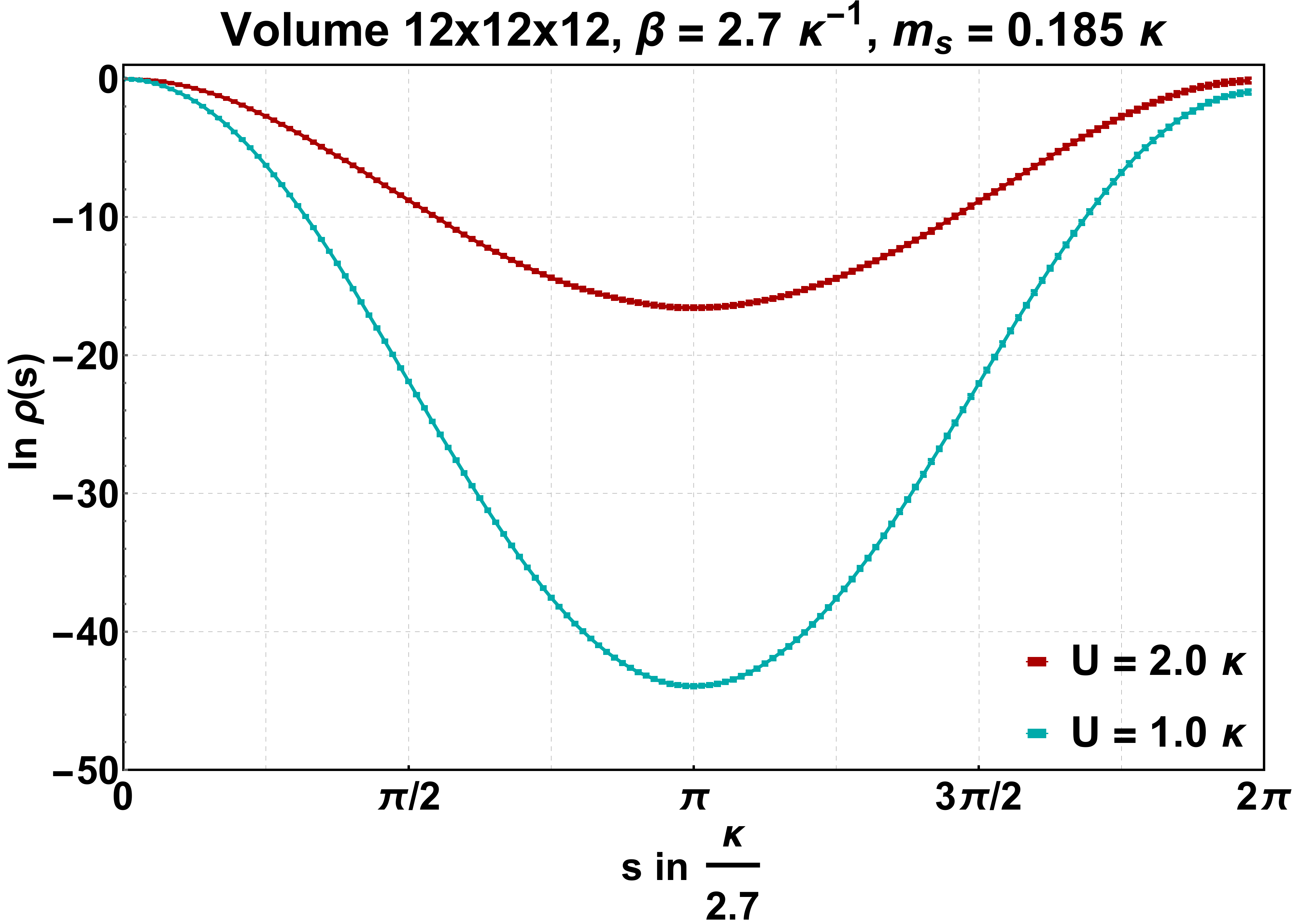}\\
 \includegraphics[width=0.45\textwidth]{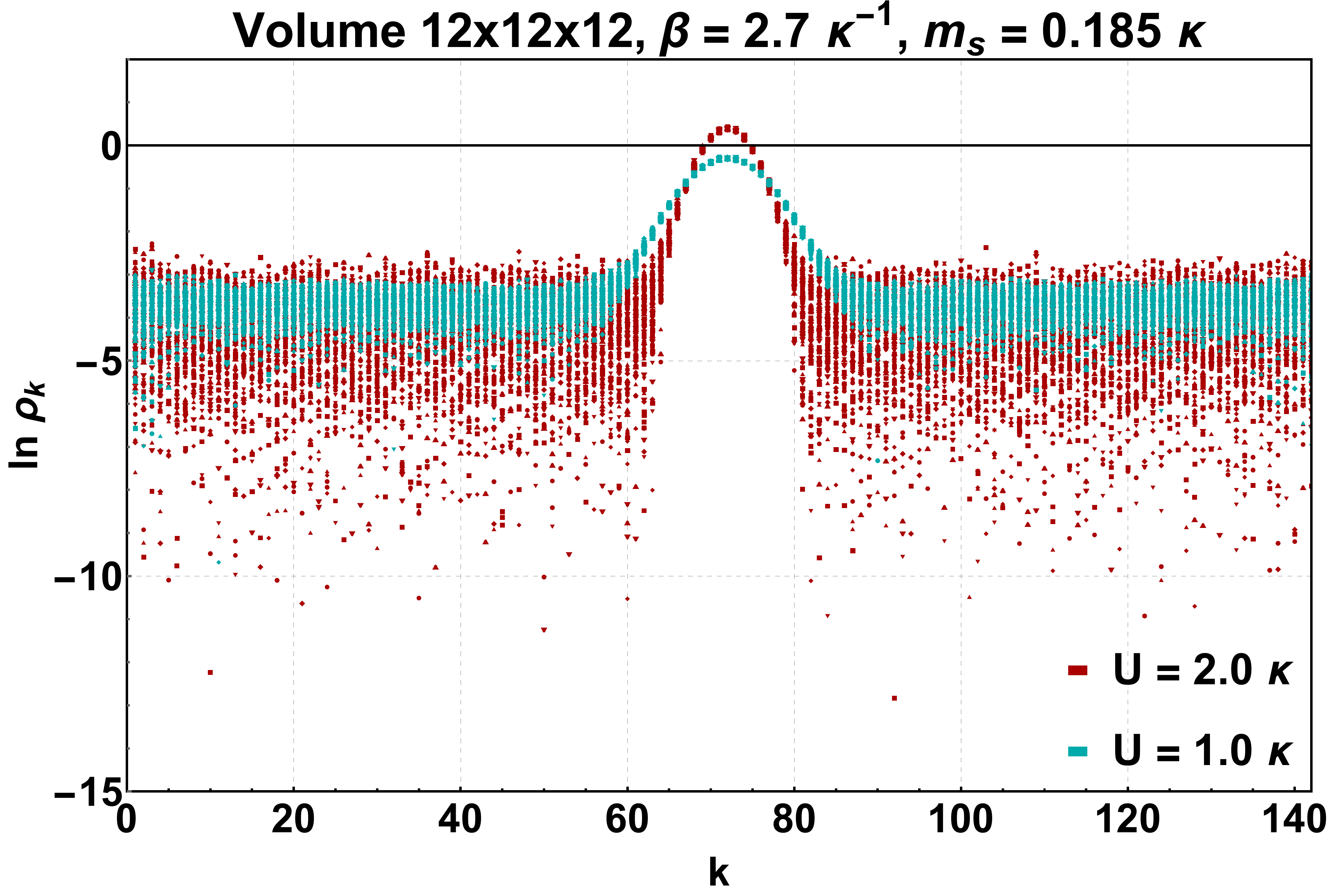}
       \caption{LLR result: $U$ dependence of $a(s)$, $\ln\rho(s)$, $\ln\tilde\rho_k$ for $N_s = N_t= 12$, 
$\beta = 2.7\, \kappa^{-1}$, $m_s=0.185 \, \kappa$. Individual bootstrap averages
are shown for $\ln\tilde\rho_k$. } \label{fig:U_dependence_vol12}
\end{figure}

\begin{figure}[htb]
       \centering
  \includegraphics[width=0.45\textwidth]{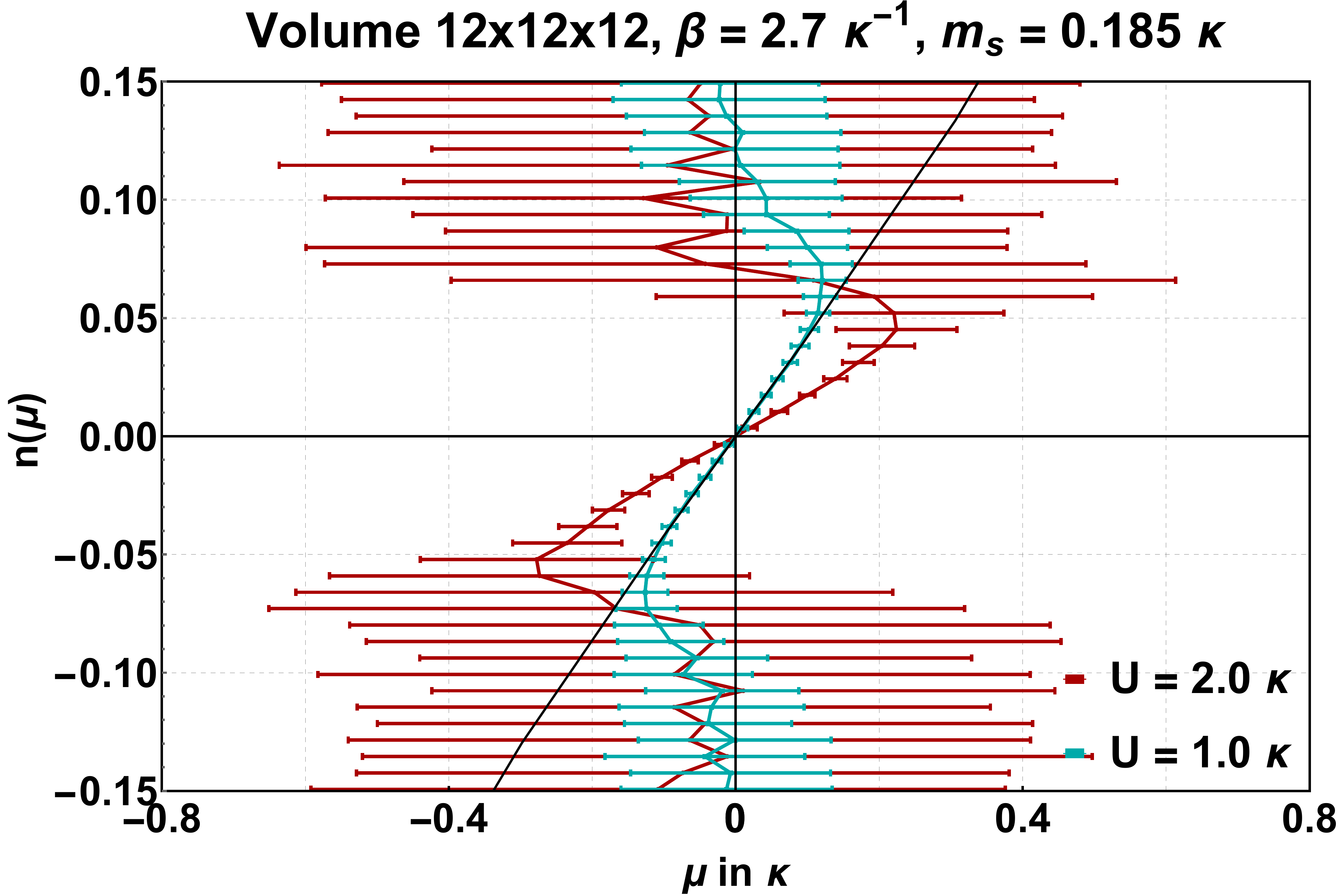}
       \caption{LLR result: $U$ dependence of $n(\mu)$ for $N_s = N_t = 12$, 
$\beta = 2.7 \, \kappa^{-1}$, $m_s=0.185 \, \kappa$. Errorbars computed by boostrapping.
Solid line shows the non-interacting tight-binding theory. 
} \label{fig:U_dependence_density_vol12}
\end{figure}

$N_s=12$ results are shown in Fig.~\ref{fig:U_dependence_vol12} for $a(s)$, $\ln\rho(s)$ and $\ln\tilde\rho_k$ and Fig.~\ref{fig:U_dependence_density_vol12} for $n(\mu)$. These confirm the qualitative changes
at $U=2.0\, \kappa$. Furthermore, a direct comparison with $N_s=6$ suggests that finite volume effects 
on $n(\mu)$ are rather mild. 

We point out here that the sign problem sets in at much smaller $\mu$ for the
larger system (as expected). While we are able to reliably compute $n(\mu)$ up to $\mu\approx 0.35 \, \kappa$ for
$N_s=6$ with $U=1.0\, \kappa$, we only reach $\mu\approx 0.1 \, \kappa$ for $N_s=12$. On the other hand, in 
both cases LLR drastically outperforms brute-force reweighting: With comparable numerical resources
we obtain a signal for the determinant ratio (\ref{eq:reweight}) up to 
$\mu\approx 0.14 \kappa$ on $N_s=6$ and $\mu\approx 0.075 \kappa$ on $N_s=12$
using the brute-force method.
While the relative advantage of LLR  becomes smaller on the larger lattice, 
we can reach much larger values of $\mu$ for $U=2.0 \, \kappa$ ($\mu\approx 0.5\, \kappa$ on $N_s=6$
and $\mu\approx 0.2 \, \kappa$ on $N_s=12$). In contrast, the $\mu$ range of reweighting is drastically
diminished at stronger coupling (cf.~Fig.~\ref{fig:signproblem2} in Sec.~\ref{sec:outlook}). 
It is this last feature which ultimately makes LLR in its present form a promising method and deserving
of further attention.

\subsection{Compressed sensing}

\begin{figure}[htb]
 \vspace{3mm}\includegraphics[width=0.45\textwidth]{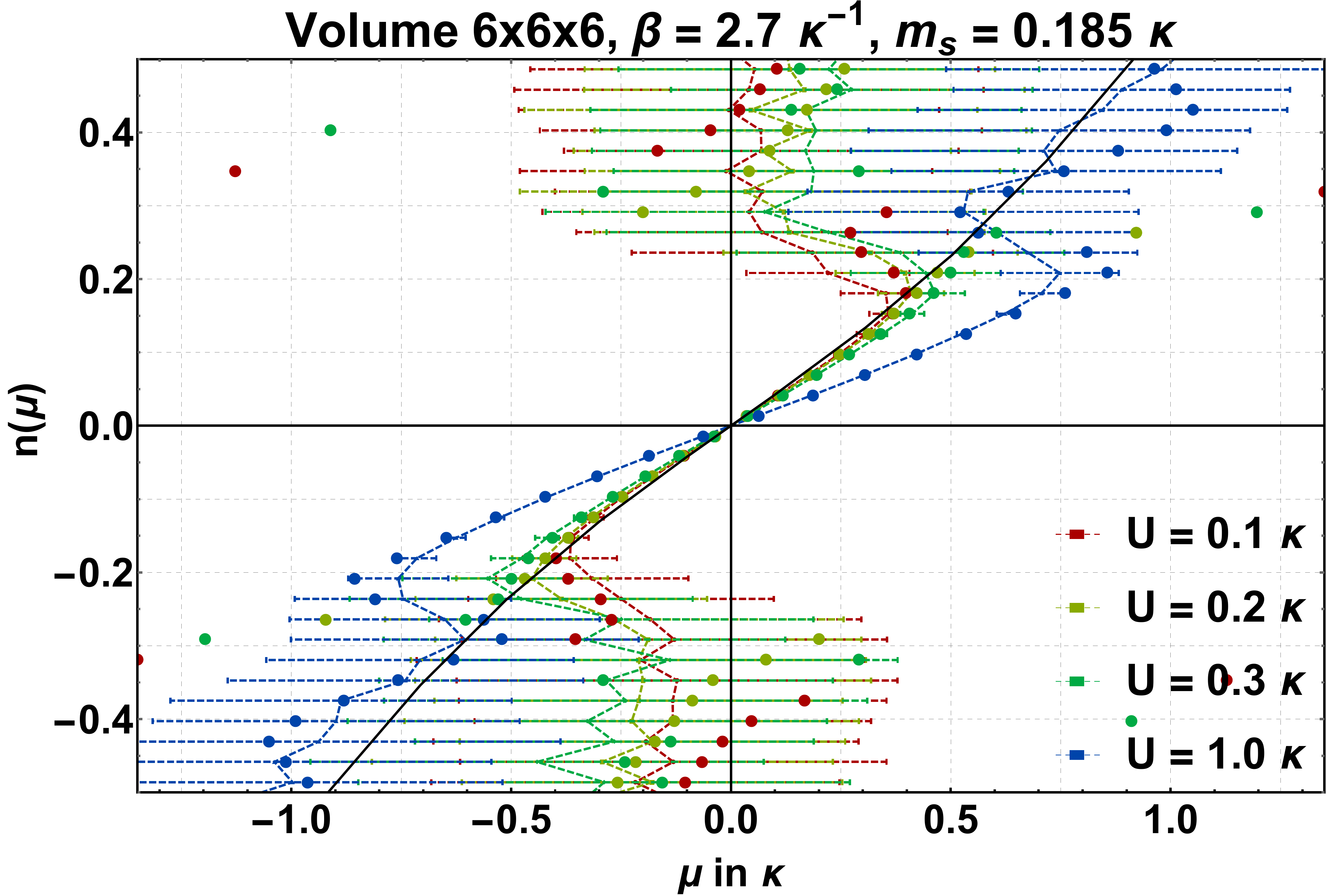}\\
 \vspace{3mm}\includegraphics[width=0.45\textwidth]{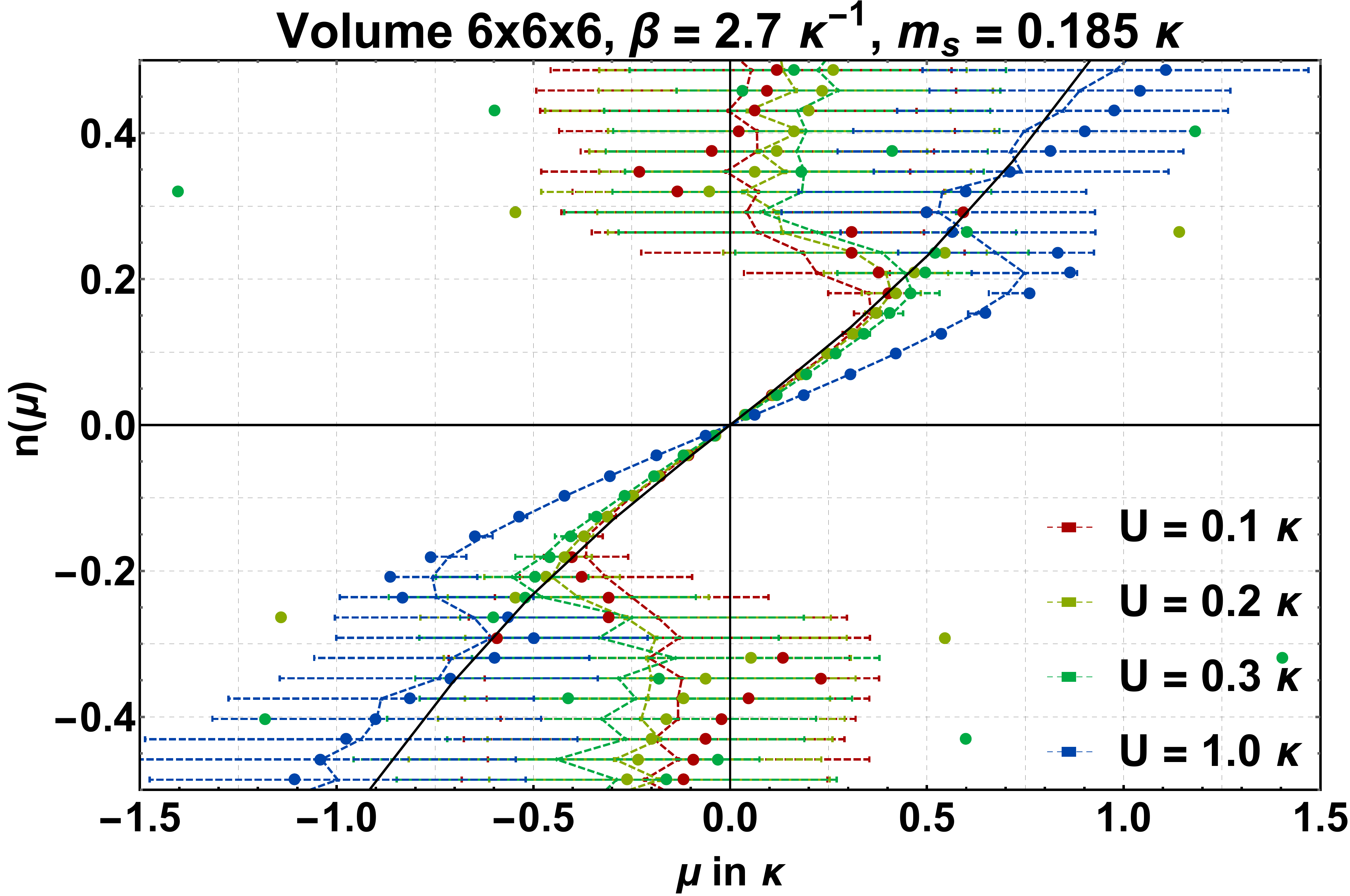}
        \caption{LLR result: $\mu$-dependence of particle density  
for $N_s = N_t = 6$, $\beta = 2.7 \, \kappa^{-1}$, $m_s = 0.185\, \kappa$ and different $U$
(results include the linear $\sim U$ term in Eq.~(\ref{eq:mu_canonical})). Dashed lines
were obtained directly from $\ln\tilde\rho_k$, while dots employed \emph{compressed sensing}:
$\ln\rho(s)$ was fit with a Fourier series (top figure) and Chebyschev polynomials (bottom figure) respectively. 
Solid line shows non-interacting tight-binding theory. }
\label{fig:compressed_sensing}
\end{figure}

Lastly, we report on our attempts to improve our results by using fit functions for $\ln\rho(s)$, a procedure referred to
as \emph{compressed sensing} in the LLR literature. The basic idea is to, instead of processing the raw data for $\ln\rho(s)$
pointwise at the supporting points $s_k$, fit the entire data set with a series expansion in some complete set of 
functions, and use the model curve to compute observables instead. The hope is that an appropriate set of functions, which
reflects the true (but a priori unknown) physics of the theory, will both suppress noise in the numerical data
for $\ln\rho(s)$ and effectively generate an interpolation to a much denser set of supporting points. This in turn
should allow for the computation of $\ln\tilde\rho_k$ at larger $k$ and hence the number density at larger $\mu$. 

Fig.~\ref{fig:compressed_sensing} shows two such attempts, where $\ln\rho(s)$ was fit with a Fourier series
and a series of Chebyshev polynomials of the first kind respectively. 
The fit function was subsequently evaluated at a much denser set of points than the original $s_k$ and used
to compute $\ln\tilde\rho_k$ and subsequently $n(\mu)$. In each case, higher order terms were added to the expansion 
until the final result stabilized. We do not obtain any errorbars. Results from the direct calculation are included 
for comparison and represented by dashed lines.

We find that these attempts do not improve the calculation of $n(\mu)$ significantly. At best, one or two additional
points (at higher densities) can be computed before the results scatter in an uncontrolled fashion. 
The Fourier series thereby seems to work only slightly better than the Chebyshev polynomials. 
We take this as an indication that additional qualitative information about the system, 
which places additional contraints on the choice of functions to use, is a necessary requirement 
for compressed sensing to be effective here. 

\section{Conclusion and outlook}\label{sec:outlook}

In this work, we have applied the Linear Logarithmic Relaxation method to the repulsive fermionic Hubbard
model on the honeycomb lattice, in order to assess its utility for alleviating the hard
sign problem of an unbalanced dynamical fermion system. A central problem thereby is the proper choice
of a target observable, which adequately reflects the complex part of the action 
and yields a generalized density of states which is suitable for further
processing. We used the average value $\Phi$ of the auxiliary (Hubbard) field to this end, which appeared
as the natural choice, as it allows for the shifting of the complex part of fermion determinant
into the bosonic sector and provides a simple integral expression (Eq.~(\ref{eq:zmu_rhos})) for obtaining the partition
function and hence the particle density. To deal with an oscillating contribution to this integral,
we chose to work in the frequency domain and divised two methods to extract the particle density from
the Fourier modes of the gDOS of $\Phi$ which essentially yields the partition function at imaginary chemical potential. Due to a slightly better performance in benchmark calculations, 
of these we chose a method based on the canonical ensembles to further process our LLR results.

We have carried out LLR calculations for a fixed temperature of $\beta=2.7\, \kappa^{-1}$, two different lattice sizes ($6^3$ and $12^3$) and different interaction strengths in the weak and intermediate coupling regime, and obtained the particle density as a function of chemical potential. We thereby observed significant deviations from
the non-interacting theory for the largest interaction strength considered, $U/\kappa=2.0$, signalling strong correlations which might eventually lead to spontaneous mass-gap formation which is known to occur at around $U/\kappa\approx 3.8$ \cite{Assaad:2013xua,Buividovich:2018crq} in the infinite volume limit.

\begin{figure}  
\begin{center}
\vspace{3mm}
\includegraphics[width=0.97\linewidth]{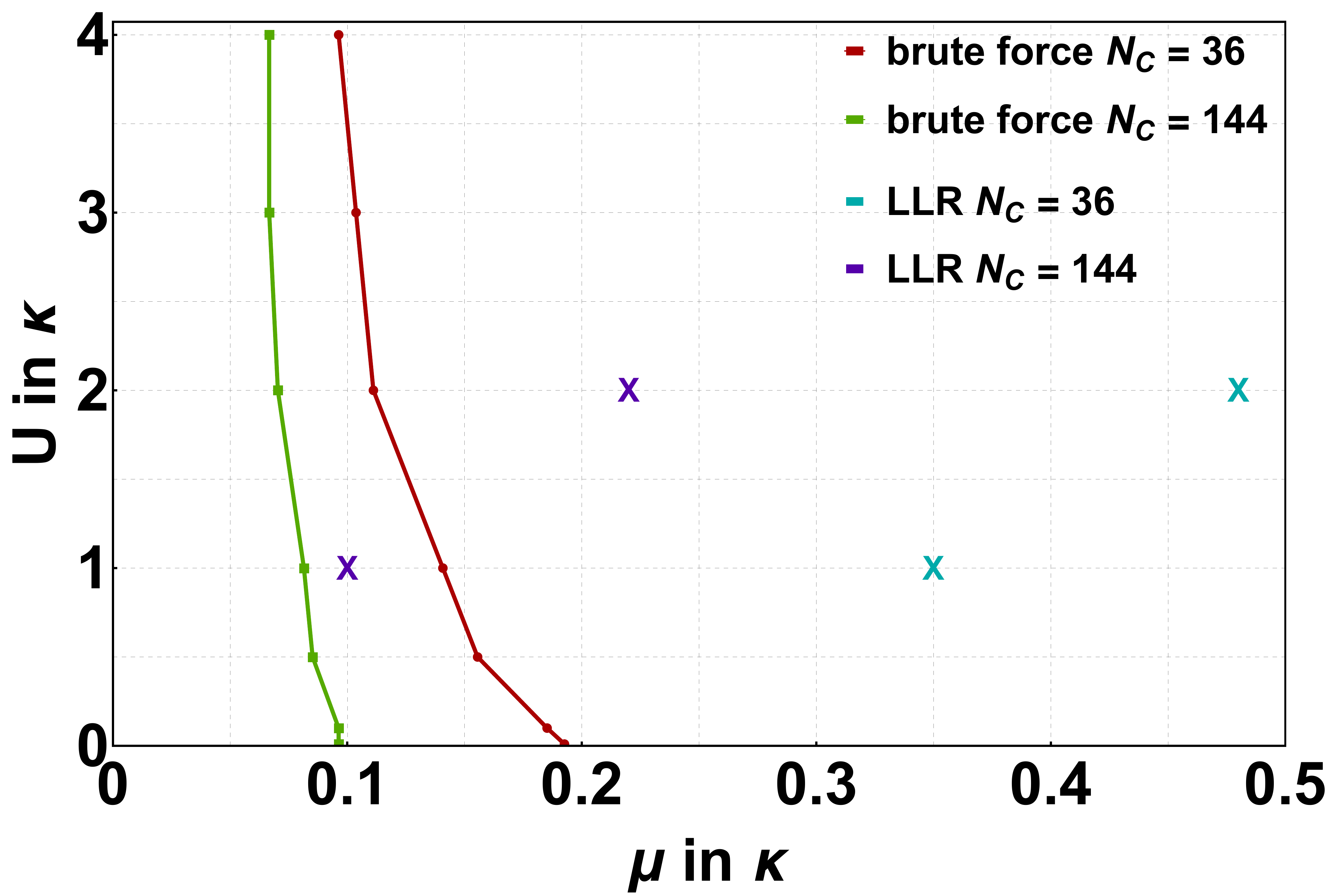}
\caption{Comparison of effective $\mu$-range of brute-force reweighting and LLR 
(results shown for $6^3$ and $12^3$ lattices at $\beta = 2.7 \, \kappa^{-1}$).
For each value of $U$ the phase distribution of $\det\widetilde{M}(\phi,\mu)/\det\widetilde{M}(\phi,-\mu)$ was measured at
different $\mu$ until the signal was lost. A roughly equal amount of computer time was 
spent for corresponding LLR calculations.   }
\label{fig:signproblem2}
\end{center}
\end{figure}

We found that using LLR in its present form, on the smaller $6^3$ lattice we are able to probe 
at least twice as far into the finite-density regime as with brute-force reweighting. While the relative advantage
of LLR is smaller on the $12^3$ lattice, we find that LLR performs much better when the interaction strength
is increased. Fig.~\ref{fig:signproblem2} shows a quantitative comparison of the effective $\mu$-ranges 
for the different parameter sets considered in this work.

Attempts to reach into higher-density regions were made
using different forms of compressed sensing, i.e.~by fitting $\ln \rho(s)$ with Fourier series and Chebyshev polynomials
and using the model curves for interpolation. While this allows us to reach slightly higher densities, we suspect
that this procedure introduces an uncontrolled systematic error, as the physics at higher densities is strongly 
sensitive to the high-frequency modes of $\rho(s)$, which such interpolations cannot account for. 

The results presented here should be taken as a proof of principle. There are several different directions
for future improvements. First and foremost, the computational resources spent for the final
calculations (i.e.~the sum of all results shown in Figs. \ref{fig:U_dependence_density_vol6} and 
\ref{fig:U_dependence_density_vol12}) were around two months of runtime
on a total of $18$ GTX $980$ Ti GPUs, which leaves much space for larger-scale projects. 
We estimate that, using the most modern hardware and libraries for sparse linear algebra, the precision
for $\ln \rho(s)$ can be increased by at least an order of magnitude. More advanced techniques for
compressed sensing could also be applied, such as Gaussian and telegraphic approximations or an
advanced moments approach, which were proposed in Ref.~\cite{Garron:2017fta}. 
Quite possibly, introducing a complex instead of a real auxiliary field has advantages, 
and in fact, it was shown that an optimal mixing factor between real and imaginary Hubbard fields exists, 
for which the sign problem is the mildest \cite{Ulybyshev:2017hbs}. LLR might also be more effective with a 
discrete Hubbard field, which is used in BSS Quantum Monte-Carlo calculations \cite{Blankenbecler:81:1,PhysRevB.34.7911}.
In addition, an alternative time-discretization with a symmetry of time reversal times sublattice 
exchange, which was proposed already in Ref.~\cite{Brower:2012zd} and recently used in
a grand canonical HMC simulation \cite{Ostmeyer:2020uov}, was shown to have strongly suppressed
discretization effects in Ref.~\cite{PhysRevB.93.155106} and might positively impact the performance of LLR as well.
And finally, there has been much recent progress regarding the Lefschetz thimble method 
\cite{Ulybyshev:2017hbs,Ulybyshev:2019fte,Ulybyshev:2019hfm}, and constructing a hybrid approach, which combines
the advantages of both methods, might be feasible. Specifically, one could attempt to apply the Lefschetz 
thimble decomposition directly to Eq.~(\ref{eq:zmu_rhos}), in order to avoid the use of reconstruction schemes
altogether and obtain a cleaner signal for $n(\mu)$. 

Taken together, we find it not unreasonable
to expect that future developments might put the van Hove singularity (VHS) of the single-particle spectrum within reach, which is of great interest in the context of superconducting phases. A crucial point thereby is the apparent stability of
the LLR technique against increases of the coupling strength $U$. Experiments on charge-doped
graphene systems have revealed a strong bandwidth renormalization 
(narrowing of the width of the $\pi$-bands) due to interactions \cite{PhysRevB.94.081403}, which suggests
that the VHS can be probed at smaller $\mu$ for larger $U$. Furthermore, a HMC study of graphene at finite
spin density revealed that the electronic Lifshitz transition at the VHS can become a true thermodynamic phase transition in the presence of interactions, with a critical temperature which increases with the coupling strength
\cite{Korner:2017qhf}. A study of an analogous transition at finite charge-carrier density thus might be feasible at large $U$, in particular as the sign problem  becomes milder at higher temperatures. 

There are many possibilities to linearize the quartic fermionic
interaction using auxiliary fields. The choice in this paper is
inspired by the observation that an explicit analytic continuation 
(i.e.~Eq.~(\ref{eq:mushift})) was sufficient  to split off the complex
part of the fermion determinant. The formulation is elegant: The
calculation of {\it one} (real) density of 
states, $\rho (s)$, is sufficient to relay the calculation of the
partiton function to one integration for each given value of the
chemical potential. Note, however, that the use of a Hubbard field
$\phi $ 
with a compact domain of support implies that the domain of the  density of states is
also compact. The calculation of such an ``intensive'' density of
states to sufficient precision is difficult~\cite{Garron:2017fta} and
the most successful LLR calculations for theories with a sign problem
are based upon non-compact densities~\cite{Langfeld:2014nta}. The use
of a non-compact formulation is left to future
work.

Lastly, we should mention that extending our work to the QCD sign
problem remains an open conceptual challenge. The system  
considered here was special since we succeeded to remove the complex
part the fermion determinant by a simple analytic continuation. 
For gauge theories no such simple transformation exists, and measuring  
a proper extensive phase is much more involved. It may well be that
this step is the most computationally demanding and contains the
central  computational complexity of the QCD sign problem.

\begin{acknowledgments}
  This work was supported by the Helmholtz International Center known as HIC for FAIR and its successor, the Helmholtz Research Academy Hessen for FAIR. 
  
We are gratefull to Pavel Buividovich, John Gracey and Maksim Ulybyshev for helpful discussions and comments.
\end{acknowledgments}

\bibliographystyle{JHEP}
\bibliography{paper-LLR}

\end{document}